\documentclass[twocolumn,twocolappendix,resetfootnote]{aastex702}

\newcommand\meerkat{J0514--4002E}
\usepackage{amsmath}

\usepackage{bm}
\usepackage{tensor}

\hypersetup{
    colorlinks=true,   
    linkcolor=blue,    
    filecolor=blue,    
    citecolor=blue,    
    urlcolor=blue,     
    pdfborder={0 0 0}  
}

\usepackage[nameinlink,noabbrev]{cleveref}
\crefname{equation}{Eq.}{Eqs.}
\Crefname{section}{Section}{Sections}
\Crefname{table}{Table}{Tables}
\Crefname{appendix}{Appendix}{Appendices}

\renewcommand{\vec}[1]{\mathbf{#1}}
\renewcommand{\d}[0]{\mathrm{d}}

\newcommand{\kmsMpc}{\,{\rm km\,s^{-1}\,Mpc^{-1}}}

\newcommand{\uhm}{Department of Physics and Astronomy, University of Hawai`i at M\=anoa, Honolulu, HI 96822, USA}
\newcommand{\asusese}{School of Earth and Space Exploration, Arizona State University, Tempe, AZ 85287, USA}
\newcommand{\asuphys}{Department of Physics, Arizona State University, Tempe, AZ 85281, USA}
\newcommand{\bu}{Department of Physics, Boston University, Boston, MA 02215, USA}
\newcommand{\iac}{Instituto Avanzado de Cosmolog\'{\i}a, Ciudad de M\'exico 10720, M\'exico}
\newcommand{\kicc}{Kavli Institute for Cosmology, University of Cambridge, Cambridge CB3 0HA, UK}
\newcommand{\ioa}{Institute of Astronomy, University of Cambridge, Cambridge CB3 0HA, UK}
\newcommand{\ifa}{Institute for Astronomy, University of Hawai`i, Honolulu, HI 96822, USA}
\newcommand{\umich}{Department of Physics, University of Michigan, Ann Arbor, MI 48109, USA}

\renewcommand{\vec}[1]{\mathbf{#1}}

\renewcommand{\d}[0]{\mathrm{d}}

\shorttitle{Cosmological coupling and BH+PSR binaries}
\shortauthors{Croker, Bowling, Cain et al.}

\begin{document}

\title{Direct inference of cosmological mass growth via timing of black hole -- pulsar binaries}

\correspondingauthor{Kevin~S.~Croker}
\email{kevin.croker@asu.edu}

\author[0000-0002-6917-0214]{Kevin~S.~Croker}
\email{kevin.croker@asu.edu}
\affiliation{\asusese}
\affiliation{\uhm}

\author[0009-0007-0782-0721]{Gibson B.\ Bowling}
\email{gbbowlin@asu.edu}
\affiliation{\asusese}

\author[0000-0001-9420-7384]{Christopher Cain}
\email{clc218@cam.ac.uk}
\affiliation{\ioa}
\affiliation{\kicc} 
\affiliation{\asusese}

\author[0000-0002-7265-7920]{Jake~Summers}
\email{jsummers@caltech.edu}
\affiliation{TAPIR, California Institute of Technology, Pasadena, CA 91125, USA}
\affiliation{LIGO Laboratory, California Institute of Technology, Pasadena, CA 91125, USA}

\author[0000-0003-1748-2010]{Duncan~Farrah}
\email{dfarrah@hawaii.edu}
\affiliation{\ifa}
\affiliation{\uhm}

\author[0000-0003-1100-9855]{Jeremy Coles}
\email{jc574@cam.ac.uk}
\affiliation{Department of Physics, Cavendish Laboratory, University of Cambridge, Cambridge CB3 0HE, UK}

\author[0000-0002-1681-0767]{Hayley Williams}
\email{hwill102@asu.edu}
\affiliation{\asusese}

\author[0000-0001-6139-649X]{Chris Pearson}
\email{chris.pearson@stfc.ac.uk}
\affiliation{RAL Space, STFC Rutherford Appleton Laboratory, Didcot, Oxfordshire OX11 0QX, UK}
\affiliation{Oxford Astrophysics, University of Oxford, Keble Rd, Oxford OX1 3RH, UK}
\affiliation{The Open University, Milton Keynes MK7 6AA, UK}

\author[0009-0004-2634-5223]{Joshua~Grumski-Flores}
\email{jgrumski@asu.edu}
\affiliation{\asuphys}

\author[0000-0003-3651-9843]{Viktor~T.~Toth}
\email{vttoth@vttoth.com}
\affiliation{Ottawa, K1N 9H5 Ontario, Canada}

\author[0009-0004-4799-891X]{Joshua~Cohon}
\email{jacohon@asu.edu}
\affiliation{\asusese}

\author[0000-0001-8156-6281]{Rogier A. Windhorst}
\email{rogier.windhorst@asu.edu}
\affiliation{\asusese}

\author[0000-0003-2911-9163]{Leander~Thiele}
\email{leander.thiele@ipmu.jp}
\affiliation{Center for Data-Driven Discovery, Kavli IPMU (WPI), UTIAS, The University of Tokyo, Kashiwa, Chiba 277-8583, Japan}
\affiliation{Kavli IPMU (WPI), UTIAS, The University of Tokyo, 5-1-5 Kashiwanoha, Kashiwa, Chiba 277-8583, Japan}

\author[0000-0003-1704-0781]{Gregory~Tarl\'e}
\email{gtarle@umich.edu}
\affiliation{\umich}

\author[0000-0002-1544-8946]{Gustavo~Niz}
\email{g.niz@ugto.mx}
\affiliation{Departamento de F\'{\i}sica, DCI-Campus Le\'{o}n, Universidad de Guanajuato, Guanajuato C.~P.~37150, M\'{e}xico}
\affiliation{\iac}

\author[0000-0001-6423-9799]{James~W.~Rohlf}
\email{rohlf@bu.edu}
\affiliation{\bu}

\author[0000-0002-3573-339X]{Nicolas Fernandez}
\email{nico.fer@rutgers.edu}
\affiliation{NHETC, Department of Physics and Astronomy, Rutgers University, Piscataway, NJ 08854, USA}
\begin{abstract}  
  Cosmologically coupled black holes (CCBHs) respond to the expansion of the Universe, independently of conventional astrophysical processes.
  Recent CCBH solutions in General Relativity demonstrate evolution of gravitating mass $m_{\rm BH}\propto a^\kappa$, where $a$ is the cosmological scale factor.
  This can be tested through the orbital period timing of black hole (BH) pulsar (PSR) binaries.
  In contrast to previous methods used to constrain $\kappa$, pulsar timing measurement baselines span decades instead of gigayears, allowing sources of bias to be well-controlled.
  As a test case, we analyze the globular cluster BH+PSR candidate J0514--4002E (spin $P\sim 5.6~{\rm ms}$, orbital period $P_{\rm b}\sim 7~\rm days$) under the assumption that the companion is a CCBH.
  Taking the difference between the observed orbital and spin period derivatives removes all kinematic effects, leaving only the pulsar’s intrinsic spin-down $\dot{P}_{\rm int}$ as unknown.
  Setting $\dot{P}_{\rm int} = 0$, its physical minimum, gives a robust upper bound $\kappa \leqslant 3.95$ at $95\%$ confidence.
  We estimate $\dot{P}_{\rm int}$ through the population of millisecond pulsars matched in spin and infer $\kappa = 0.82 \pm 1.67$ and $0.75 \pm 1.54$, using $H_0 = 67.6$ and $73.5\kmsMpc$, respectively.
  Adopting typical baseline scaling of timing precision, the value $\kappa = 3$, argued elsewhere to link energized vacuum BH interiors to accelerated cosmological expansion, can be excluded or detected for this $2.35^{+0.20}_{-0.18}\,M_\odot$ companion at $3\sigma$ and $5\sigma$ confidence by 2031 and 2036, respectively.
\end{abstract}
\section{Introduction}
Binary pulsars continue to provide the most precise tests of relativistic orbital dynamics \citep[e.g.][]{TaylorFowler79,KramerStairs21}.
Predictions for these astrophysical systems are typically determined in the point-mass limit of post-Newtonian (PN) frameworks, expansions of gravitational dynamics in dimensionless velocity squared $(v/c)^2$ \cite[e.g.][]{DamourTaylor92,TEGP18}.
Timing measurements for binary systems can be done generically within metric theories of gravity, allowing all manner of novel General Relativity (GR) predictions, and possible departures from GR \citep[e.g.][]{YunesSiemens25}, to be interrogated.

Cosmological coupling (CC) is a recent prediction in GR that highly localized regions of relativistic stress can (not must) influence global cosmological dynamics at leading order \cite[e.g.][]{CrokerWeiner19,CrokerWeiner22}.
The effect becomes relevant given the existence of non-singular and horizon-free BH solutions in GR \citep[e.g.][]{Dymnikova02,MazurMottola15}, first posited by \citet{Gliner66} to form as regions of energized vacuum during stellar gravitational collapse.
If BHs are non-singular and cosmologically coupled, they could drive accelerated cosmic expansion by transmuting baryons at stellar collapse into energized vacuum \citep[e.g.][]{Farrah23b,CrokerTarle24}.
\citet{Ahlen25} have shown that the resulting cosmology mimics the primary observables of $\Lambda$CDM, yet halves the Hubble tension and recovers a positive summed neutrino mass consistent with terrestrial and Lyman-$\alpha$ constraints \citep{Esteban24,CainEngelen25,Elbers25,Kageura26}.

The reciprocal effect, whether and how cosmologically coupled stress responds locally to cosmological dynamics, is a more nuanced question.
In contrast to linear theories like electrodynamics, the existence of a solution in GR that satisfies cosmological boundary conditions does not guarantee uniqueness of that solution \citep{BenderOrszag78}.
Recent progress has instead focused on large classes of solutions.
\citet{FaraoniRinaldi24,FaraoniRinaldi26} have established via geometric methods that the radii of BH apparant horizons must evolve toward comovement within a Robertson-Walker (RW) cosmology.
Because radius often relates to gravitating mass in compact object solutions, a natural local CC phenomenology is to adjust the gravitating mass $m \propto a^\kappa$ so that averages over a population of CCd systems reproduce large-scale CC phenomenology \citep{CrokerZevin21}.
The existence of $\kappa := 0$ solutions, like those of \citet{Kottler18,McVittie33,EinsteinStrauss45,GaurVisser24}, does not exclude $\kappa \neq 0$%
\footnote{The most well-known cosmological BH, that of \citet{McVittie33}, is engineered to be cosmologically unresponsive.
It breaks in the way highlighted by \citet{FaraoniRinaldi24}: a naked null singularity exists at its apparant horizon.
The well-known Schwarzschild-deSitter solution found by \citet{Kottler18} is fully static.
It is too algebraically special to be observationally relevant.
Any astrophysically relevant BH model must accommodate arbitrary RW asymptotics, because BHs persist through matter domination and the transition toward DE domination.
If considering primordial BHs, then BHs additionally persist through radiation domination and the transition into matter domination.
}.
\citet{CadoniSanna23,CadoniMurgia24,CadoniPitzalis24,CadoniPitzalis25,CadoniDelima26} have constructed local solutions for BHs with $\kappa = 1$.
Moreover, \citet{CrokerCohon26} have constructed a point-mass model in linearized GR, where the Hernandez-Misner-Sharp (HMS) mass evolves with any $|\kappa| \leqslant 3$ permitted. 
Because the HMS mass is geometrical, orbital dynamics would respond observably to this growth.

Observational searches for purely cosmological mass growth are complicated by the slowness of cosmological effects over most of the universe's history.
Existing approaches can be broadly divided into three categories: differential population studies, BH systems with age proxies, and aggregate population studies.
Differential studies of the SMBH populations of quiescent elliptical galaxies by \cite{Farrah23a, Farrah23b, Farrah25}, which carefully account for measurement and selection bias, find consistency with $\kappa \sim 3$ passive cosmological growth at $z \lesssim 1.8$.
Inspiraling SMBHs source the low frequency portion of the stochastic gravitational wave background and the anomalously large signal amplitude reported by NANOGrav~\citep{NANOgravFirst} could be a signal of $\kappa \sim 3$ growth in this population \citep{Lacy23, CalzaFrancesco25}.

For lower-mass BHs, the situation is less clear.
Stellar mass BHs in environments where reliable ages can be estimated, as well as the LIGO merging population, can be used to constrain $\kappa$ once a minimum BH mass is asserted.
Adopting $m_{\rm BH} \geqslant 2\,M_\odot$ gives $\kappa \lesssim 0.5$ from all these populations \citep{CrokerZevin21,Rodriguez23,Ghodla23,AndraeElbadry23,mlinarzwitter24} except possibly low-mass X-ray binaries \citep{GaoLi2023}.
\cite{Amendola24} have pointed out that a lower-bound BH mass from stellar-collapse is a feature of Kerr-like singular solutions, however, and need not hold in general for non-singular BHs.
Even so, $\kappa \sim 3$ would imply birth masses $\sim 0.1 M_\odot$ are possible \citep[although not mandatory, e.g.][]{MillerJones21}.
Finally, \citet{Shah25} find no increased scatter in supernovae magnitudes at $z \lesssim 1$ due to strong gravitational lensing.
Such a signature is expected from a uniformly distributed cosmologically coupled BH population abundant enough to drive accelerated expansion, but a consistent confrontation with data where CCBH lensing strength decreases backward in time $\propto a^\kappa$ has yet to be performed.
To begin to make sense of this observational landscape, we require a direct probe of the phenomenon.

In this paper, we utilize the exquisite timing precision of pulsar binaries to perform the first direct inference of $\kappa$ with the BH+PSR candidate \meerkat~reported by \citet{meerkat24}.
From \meerkat, we extract a robust observational upper-bound for $\kappa$.
We then leverage the larger Galactic field PSR population to present the first direct inference of $\kappa$.
Given the existing baseline data, we highlight that $\kappa \sim 3$ can be confirmed or refuted in \meerkat~at $3\sigma$ confidence within the next four years.
In a comprehensive Appendix \ref{sec:theory}, we generalize well-known results on PN orbital mechanics to the case of a single time-varying mass with a preferred frame (the Cosmic Microwave Background, CMB) via Hamiltonian methods and the Poisson bracket.
For dimensional clarity, we explicitly indicate light speed $c$.

\section{Methods}
Using data from the Green Bank Telescope and the Karoo Array Telescope (MeerKAT), \citet{meerkat24} report timing observations of \meerkat, a $5.6\,\rm ms$ PSR in an eccentric binary orbit within GC NGC 1851.
The system is noteworthy because the pulsar's unseen companion's mass is $2.09$ to $2.71\,M_\odot$ at 95\% confidence, placing the companion in the mass gap between known neutron stars and BHs.
In \cref{tbl:meerkat}, we provide the relevant orbital properties of J0514--4002E as determined by \citet{meerkat24} via a global post-Keplerian fit.
In \S\ref{sec:CC_periastron}-\S\ref{sec:CC_orbital_i} we show that CC does not measurably impact the periastron precession $\dot{\omega}$, Einstein delay $\gamma_{\rm E}$, and orthometric Shapiro amplitude $h_3$ used to extract these mass measurements from \meerkat.
\begin{deluxetable}{lll}
  \caption{PSR J0514--4002E Binary Orbital and Pulsar Parameters by \citet{meerkat24}\label{tbl:meerkat}.
  Reported uncertainties are $68\%$ confidence.
  Dots denote precision beyond what is required for computation in this work.}
  \tablehead{ \colhead{Parameter} & \colhead{Value} & \colhead{Note}} 
  \startdata
  $M$ & $3.8870 \pm 0.0045 ~\rm M_\odot$ & Total mass \\
  $m_{\rm p}$ & $1.53^{+0.18}_{-0.20}~\rm M_\odot$ & Pulsar mass \\
  $P_{\rm b, obs}$ & $6.43\ldots\times10^{5}$~s & \\ 
  $\dot{P}_{\rm b,obs}$ & $(18.1 \pm 5.6)\times 10^{-12}$ & $\rm s\,s^{-1}$ units \\
  $P_{\rm obs}$ & $0.005595\ldots$~s & \\
  $\dot{P}_{\rm obs}$ & $(1.9330 \pm 0.0013)\times 10^{-19}$ & $\rm s\,s^{-1}$ units \\
  $e$ & $0.70793232 \pm \ldots$ & \\
  $T_\star$ & $53731$ MJD $\to$ $59793$ MJD &
  \enddata
\end{deluxetable}%

\citet{Freire17} gives the relation between observed $\dot{P}_{\rm b,obs}$ and intrinsic $\dot{P}_{\rm b,int}$ orbital period derivatives for pulsar binaries in globular clusters (GCs),
\begin{align}
  \frac{\dot{P}_{\rm b,obs}}{P_{\rm b}} = \frac{\dot{P}_{\rm b,int}}{P_{\rm b}} + \frac{\mu_{\rm p}^2d}{c} + \frac{a_{\ell,\rm GC}}{c} + \frac{a_{\rm gal}}{c}. \label{eqn:period_variation}
\end{align}
Here $\mu_{\rm p}$ is the composite proper motion, $d$ is the distance to the pulsar binary center-of-momentum (the Shklovskii term, often approximated via the distance to the GC center), $a_{\rm\ell,GC}$ is the line-of-sight acceleration due to the immediate gravitational environment of the system, and $a_{\rm gal}$ is the acceleration difference between the Solar system and the target GC within the galaxy field, projected along the direction to the cluster.
\meerkat~shows no evidence of material in the binary environment, so there is no orbit evolution from accretion or mass-loss.

We assume the companion is a cosmologically coupled BH, with far field physics as modeled by \citet{CrokerCohon26}.
In \S\ref{sec:CC_period_change}, we show that the only additional measurable GR contribution (by 6 orders of magnitude) to orbital period evolution in BH+PSR binaries due to CC is,
\begin{align}
  \left<\frac{\dot{P}_{\rm b}^{\rm CC}}{P_{\rm b}}\right> = -\kappa H\left[\left(3 - \frac{1}{1+q}\right) + \mathcal{O}(\epsilon)\right],
\end{align}
per orbit.%
\footnote{
Contrast with the more familiar ``Jeans type'' mass accretion/loss, given by \citet[][\S IV, Eq.~4.1]{DamourTaylor91} as $\dot{P}_{\rm b}^{\rm mass}/P_{\rm b} = -2\kappa H/(1+q)$.
In the \citet{CrokerCohon26} model, the presence of ambient density and a lack of time-translation symmetry in RW cosmology support conductive flow (not accretion) onto the central object.
This conserves total angular momentum, instead of the specific angular momentum conserved in Jeans-type mass change.}
Here we have introduced the mass ratio,
\begin{align}
  q := m_{\rm p}/m_{\rm c},
\end{align}
where $m_{\rm p}$ and $m_{\rm c}$ are the instantaneous pulsar and companion (CCBH) masses, respectively, and $\epsilon \lesssim 10^{-6}$ is defined in \cref{eqn:epsilon}.

As shown in \S\ref{sec:higher_PN}, orbital period change from CC complements that from radiative losses, so that intrinsic orbital period evolution becomes,
\begin{align}
  \frac{\dot{P}_{\rm b,int}}{P_{\rm b}} = \frac{\dot{P}_{\rm b}^{\rm CC}}{P_{\rm b}} + \frac{\dot{P}_{\rm b}^{\rm GW}}{P_{\rm b}}. \label{eqn:defn_b_int}
\end{align}
We assume that no additional radiative degrees of freedom are at play.

Standard analyses assume $\kappa := 0$, and thus $\dot{P}_{\rm b}^{\rm CC} = 0$.
Because Shklovskii $\mu_{\rm p}^2d$ and Galactic $a_{\rm gal}$ accelerations are typically small, \cref{eqn:period_variation,eqn:defn_b_int} imply that any observed $\dot{P}_{\rm b,obs} \gg \dot{P}_{\rm b}^{\rm GW}$ is naturally interpreted as an estimate of $a_{\ell,\rm GC}$.
Because pulsar spin derivative $\dot{P}_{\rm obs}$ and orbital period derivative $\dot{P}_{\rm b,obs}$ are both offset by the same secular Doppler shift, the kinematic contributions to $\dot{P}_{\rm obs}$ are identical to those in \cref{eqn:period_variation},
\begin{align}
  \frac{\dot{P}_{\rm obs}}{P} = \frac{\dot{P}_{\rm int}}{P} + \frac{\mu_{\rm p}^2d}{c} + \frac{a_{\ell,\rm GC}}{c} + \frac{a_{\rm gal}}{c}. \label{eqn:spin_variation}
\end{align}
An orbital estimate of $a_{\ell,\rm GC}$ can then be used to compute the intrinsic pulsar spin derivative $\dot{P}_{\rm int}$, if the Shklovskii and Galactic terms are known.

If the BH is cosmologically coupled with unknown $\kappa$, however, $a_{\ell,\rm GC}$ cannot be extracted from the orbital timing of the GC binary because it is degenerate with $\dot{P}_{\rm b}^{\rm CC}$ in \cref{eqn:period_variation,eqn:defn_b_int}.
For the same reason, inference of $\dot{P}_{\rm b}^{\rm CC}$, and thus $\kappa$, cannot be performed from orbital timing alone.
We can resolve this degeneracy by inverting the traditional procedure and using the spin period derivative as a proxy for kinematic effects. Formally, we subtract \cref{eqn:spin_variation} from \cref{eqn:period_variation,eqn:defn_b_int} then rearrange to find,
\begin{align}
  \frac{\dot{P}_{\rm b}^{\rm CC}}{P_{\rm b}} = \frac{\dot{P}_{\rm b,obs}}{P_{\rm b}} - \frac{\dot{P}_{\rm obs}}{P} - \frac{\dot{P}_{\rm b}^{\rm GW}}{P_{\rm b}} + \frac{\dot{P}_{\rm int}}{P}. \label{eqn:difference_trick}
\end{align}
This relation is entirely between binary observables and quantities intrinsic to the binary orbit and the bound objects.
In principle then, if $\dot{P}_{\rm int}$ can be constrained through other data, the observed derivative in spin period $\dot{P}_{\rm obs}$ can be used to estimate the intrinsic derivative in orbital period $\dot{P}_{\rm b}^{\rm CC} + \dot{P}_{\rm b}^{\rm GW}$ with exactly zero kinematic contamination.

\subsection{Calibrator Pulsar Selection}
\begin{figure}[t!]
    \centering
    \includegraphics[width=\linewidth]{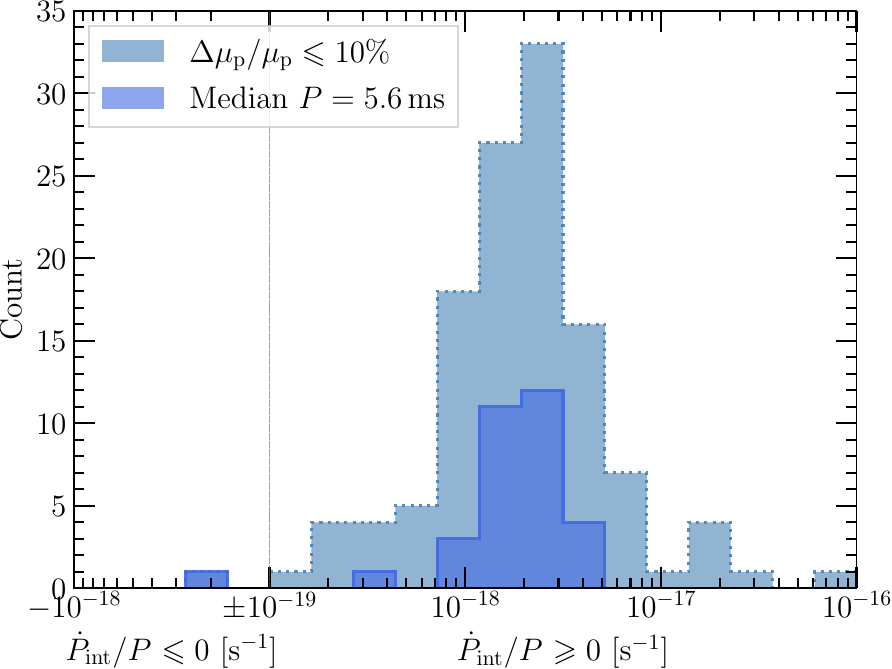}
    \caption{
      Distributions of $\dot{P}_{\rm int}/P$ from $1 \leqslant P \leqslant 10$ millisecond pulsars, selected from the ATNF catalog to be free of local accelerations.
      (Light blue, dotted) Field millisecond pulsars with proper motion measured to fractionally better than $10\%$ and, if in a binary, have compact companion.
      (Dark blue, solid) The subsample $(N=32)$ with median period matched to \meerkat, $P = 5.6\,\rm ms$.
      This sample is adopted as a prior for the intrinsic period derivative $\dot{P}_{\rm int}$ of \meerkat.
    }
    \label{fig:prior}
\end{figure}%
To determine an empirical prior on $\dot{P}_{\rm int}/P$, the quantity which enters \cref{eqn:difference_trick}, we begin with all millisecond pulsars ($1 \leqslant P \leqslant 10\,\rm ms$) from the Australia Telescope National Facility (ATNF) pulsar catalog%
\footnote{\url{https://www.atnf.csiro.au/research/pulsar/psrcat/}}.
We select only Galactic field pulsars so that $a_{\ell,\rm GC} = 0$ in \cref{eqn:spin_variation}.
Field pulsars are identified using the ATNF GC associations and the catalog of known pulsars in GCs\footnote{\url{https://www3.mpifr-bonn.mpg.de/staff/pfreire/GCpsr.html}} maintained by P.~C.~C. Freire, which is based on the McMaster catalog of Galactic globular clusters \citep{Harris1996, Harris2010}.
For field pulsars within binary orbits, we select systems with compact companions that cannot contribute mass outflows.
Thus, there is no Jeans-type accretion contaminant.
We then require fractional uncertainty in proper motion $\Delta \mu_{\rm p}/\mu_{\rm p} \leqslant 10\%$, to enable accurate estimation of $\mu_{\rm p}^2 d$ and $a_{\rm gal}$ within \cref{eqn:spin_variation}.
For each object, the acceleration difference $a_{\rm gal}$ is computed using the \citet{McMillan2017} Galactic model and the Shklovskii acceleration $\mu_{\rm p}^2 d$ is computed using the proper motions and distances%
\footnote{Distances were taken directly from ATNF's \texttt{DIST} parameter to maintain a consistent distance prescription across the sample} 
tabulated in the catalog.
Combined with the reported $\dot{P}_{\rm obs}$ and $P$, \cref{eqn:spin_variation} is used to compute $\dot{P}_{\rm int}/P$.

The resulting sample of $N=124$ pulsars has $\dot{P}_{\rm int}/P = 2.0^{+2.5}_{-1.8}\times 10^{-18}\,\rm s^{-1}$ at $68\%$ confidence and is shown in \cref{fig:prior}, light-blue.
We next construct the sample matched to \meerkat~in median period $P=5.6\,\rm ms$, giving $N=32$ pulsars.
This sample has $\dot{P}_{\rm int}/P = 2.0^{+0.9}_{-0.8}\times 10^{-18}\,\rm s^{-1}$ at $68\%$ confidence and is shown in \cref{fig:prior}, medium-blue.
This sample determines our empirical prior on $\dot{P}_{\rm int}/P$.

The intrinsic spin derivatives computed for all calibrator pulsars are determined independently of any orbit.
Thus, they are not affected by the presence or absence of CC.
We exclude millisecond PSRs within GCs from the calibrator sample because estimation of $\dot{P}_{\rm int}$ from this population requires orbital timing.
The $a_{\ell,\rm GC}$ inferred from $\dot{P}_{\rm b,obs}$ would be contaminated by $\dot{P}_{\rm b}^{\rm CC}$ if the companion were actually a CCBH with $m_{\rm BH} < 2\,M_\odot$.
Inclusion of contaminated systems would bias $\kappa$ away from its true value toward zero.
GC pulsars with optically or spectroscopically confirmed white dwarf companions avoid this ambiguity, but we defer their inclusion to future work.

\section{Results}
\label{sec:results}
\begin{figure}[t!]
  \centering
  \includegraphics[width=\linewidth]{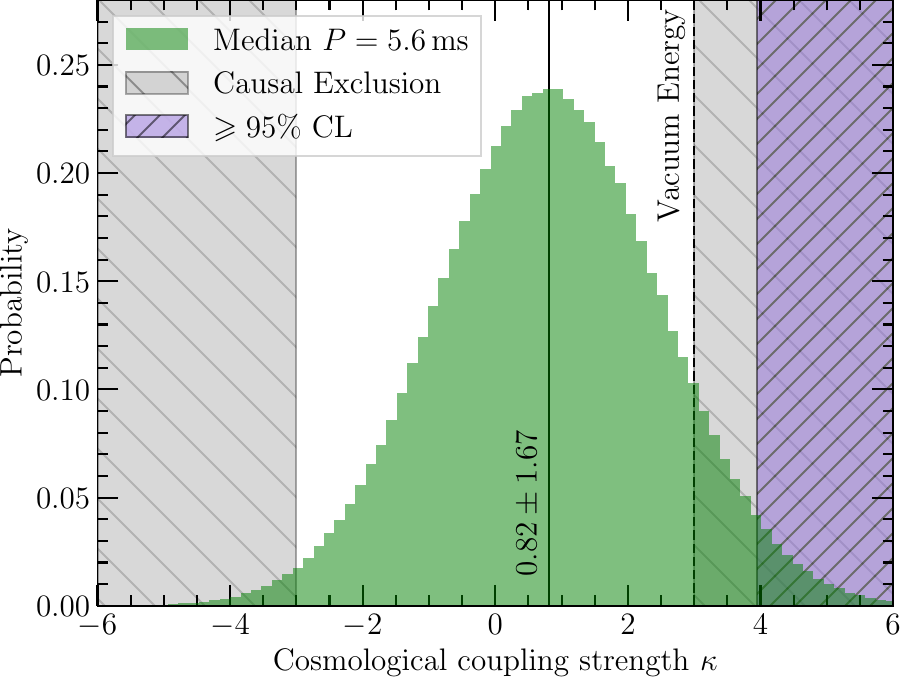} 
  \caption{\label{fig:kappa}
    Constraints on stellar-mass BH $\kappa$ from BH+Pulsar candidate \meerkat, assuming \emph{Planck} $H_0$ \citep{PlanckPR4}.
    Orbital period derivative, and thus $\kappa$, is determined for this globular cluster pulsar via the intrinsic spin prior matched in median pulse period to $5.6\,\rm ms$, displayed in \cref{fig:prior}.
    Values for $\kappa$ that classically imply causality violation (grey, hatched) are indicated.
    Robust exclusion at $\geqslant 95\%$ confidence (purple) is independent of intrinsic spin.
  }
\end{figure}%
Any pulsar without evidence of ongoing mass accretion (active recycling) must physically satisfy $\dot{P}_{\rm int} > 0$.
From \cref{eqn:decay_cons,eqn:difference_trick}, any physical $\dot{P}_{\rm int} > 0$ decreases the inferred $\kappa$, because both terms contribute to $\dot{P}_{\rm b,obs}$ with the same sign.
Setting $\dot{P}_{\rm int} := 0$ and adopting \emph{Planck} PR4 $H_0$ from \citet{PlanckPR4} then gives a robust upper bound%
\footnote{The \emph{Planck} PR4 $H_0$ is smaller than that measured by \citet{Casertano26}, and so provides the more conservative upper-limit}
on $\kappa$, 
\begin{align}
  \kappa \leqslant 3.95 \qquad 95\%~\rm CL. \label{eqn:kappa_upper_bound}
\end{align}
We display this exclusion in \cref{fig:kappa}.
We next compute $\kappa$ with \cref{eqn:decay_cons,eqn:peters_dotPb} in $N=10^6$ draws from the parameters in \cref{tbl:meerkat} regarded as gaussian (we symmetrize $m_{\rm p}$ to $\pm 0.20$), and from the median-matched sample in \cref{fig:prior} with replacements, to find
\begin{align}
  \kappa =
  \begin{cases}
    0.82 \pm 1.67 & (\text{\emph{Planck} PR4}) \\
    0.75 \pm 1.54 & (\text{H0DN})
  \end{cases}, \label{eqn:kappa_inference}
\end{align}
where we also include $H_0$ from the most recent measurement by the \citet{Casertano26}.
Both values are consistent with existing constraints on $\kappa$.
The Dominant Energy Condition (DEC) requires $-3 \leqslant \kappa \leqslant 3$ for stressed causal material.
\begin{deluxetable}{ccll}
  \caption{Timing baseline endpoints by which \meerkat~alone excludes or detects literature-predicted $\kappa$.
    Assuming $T^{-2.3}$ improvement in orbital period derivative precision with a timing baseline anchored in 2005, we report endpoints for $50\%$ and $95\%$ probability of achieving the target confidence.
    Limitations based on our current $N=32$ cleaned and median-matched field millisecond pulsar calibrator for $\Delta\dot{P}_{\rm int}$ indicated with $\infty$.
    Median endpoint reported if uncertainty spans months, $68\%$ confidence interval reported if uncertainty spans years (see \S\ref{sec:forecast}).
    \label{tbl:dakine}}
  \tablehead{\colhead{\shortstack{Target \\ $\kappa$}} & \colhead{\shortstack{Exclude /\\ Detect}} & \colhead{\shortstack{Baseline End \\ ($50\%$)}} & \colhead{\shortstack{Baseline End \\ ($95\%$)}}}
  \startdata
  1 & $2\sigma$ & January 2035 & 2046--2051 \\
  1 & $3\sigma$ & 2041--2043 & 2058--$\infty$ \\
  1 & $5\sigma$ & $\infty$ & $\infty$ \\
  \hline
  3 & $2\sigma$ & May 2023\tablenotemark{a} & October 2028 \\
  3 & $3\sigma$ & November 2026\tablenotemark{b} & July 2031 \\
  3 & $5\sigma$ & June 2032 & August 2036
  \enddata
  \tablenotetext{a}{Baseline of \citet{meerkat24} ends in August 2022}
  \tablenotetext{b}{To our knowledge, the most recent (proprietary) timing data are from January 2026.
  Our forecast motivates increased cadence.}
\end{deluxetable}%
\cref{eqn:kappa_upper_bound,eqn:kappa_inference} do not yet discriminate any physical value of $\kappa$ from 0.
Instead, they are a consistency test on the method: there is no \emph{a priori} constraint on $\kappa$ and the method could have recovered values entirely incompatible with the DEC.

Radio pulsar timing uncertainties scale reliably with baseline observation.
In \S\ref{sec:forecast}, we compute the additional baseline required for a desired confident detection of some $\kappa_{\rm true}$ at some desired significance.
The additional baselines computed from \cref{eqn:forecast_baseline}, assuming detector cadence about halfway between continual and endpoint, are given in \cref{tbl:dakine}.

\section{Discussion}
Our prior on the intrinsic period derivative, shown in \cref{fig:prior}, could be improved through Bayesian population inference to include selection effects.
This becomes more important once $3\sigma$ exclusion or detection of $\kappa$ becomes possible for \meerkat~(c.f., \cref{tbl:dakine}), and the uncertainty in the parameters of J0514--4002E begin to shrink closer to that of our currently adopted prior's width.
In this study, we have taken the prior to be a discrete set of observed period derivatives, while, in general, one should use the observed values to characterize the underlying population distribution for the intrinsic period derivative \citep[e.g.][]{Alsing2017}.
For example, one could assume that the true prior is Gaussian in $\log(\dot{P}_{\rm int}/P)$ with some mean and variance, then infer the mean and variance from the observed values.
Doing so would incorporate the uncertainties from each millisecond pulsar system, as well as selection effects.
By this, we mean that some portions of our sample of field millisecond pulsars could be over-represented \citep[e.g.][]{Lorimer08, Drummond26}.
Working with the inferred underlying distribution of pulsar intrinsic period derivatives will enable more robust inference of $\kappa$ in future work.

The Square kilometer Array Observatory (SKAO) offers exciting prospects for pulsar detection and timing precision in Phase 1 of the SKAO using two telescopes with SKA-Low (50-350 MHz) and SKA-Mid (350 MHz-15.4 GHz) initially with the AA* array assembly (SKA-Low 307 stations 78,592 antennae from 2030, SKA-Mid 144 dishes from 2032) evolving to the full AA4 array assembly (SKA-Low 512 stations 131,072 antennae, SKA-Mid 197 dishes). 
In light of the MeerKAT experience, models need updating, but based on population studies and models in \citet{Keane25}, the AA* array assembly is predicted to detect 820 millisecond pulsars (440 with SKA-Low, 380 SKA-Mid) within 5 degrees of the Galactic plane, rising to over 1000 predicted for AA4 array assembly.
This population has the potential to significantly strengthen the calibrator sample.

In addition, approximately 110 (AA*) to 140 (AA4) binary neutron star systems are predicted.
SKA1-Low detections will be mostly restricted to the local sky due to luminosity and interstellar medium effects, plus those more distant (out to 10-15 kpc bringing local group galaxies into the frame) at higher Galactic latitudes.
Pulsars in the latter group will likely be young with high luminosity and correlate with supernova remnants. 
Since the SKA will be able to perform acceleration searches (up to and including $350~\mathrm{m}~\mathrm{s}^{-2}$) on all of the incoming pulsar search data, detection of possible BH+PSR systems in the Galactic disk becomes possible where the BH mass is at least $15~\rm M_\odot$ \citep{SKAprospects}.

In order to translate the above considerations into an estimate of how many BH+PSR binaries the SKA may observe, we need a model for the population of such objects in the Galactic field.
One such model that focuses on the formation of BH+PSR systems from isolated massive binary stars uses the binary population evolution synthesis code COMPAS and predicts the presence of 50--2000 binaries.
Using mock-surveys, such as those done by \cite{MeerKAT09} for MeerKAT and assumptions about the survey parameters (in particular the area of the sky surveyed and the integration time), \citet{SKAmodelling} predict that ``the SKA will observe 1--80 PSR+BHs, with 0--4 binaries containing millisecond pulsars.''
The uncertainty comes from feasible pulsar evolution, massive binary evolution, and in re-scaling their model to the Milky Way population (e.g. assumptions about the number of stars in the Milky Way, the binary fraction, and initial mass function). 

In summary, at coupling strength $\kappa>0$, BH+PSR binaries receive an additional contribution to the rate of orbital period change $\propto -\kappa H_0P_{\rm b}$.
Orbital period timing of BH+PSR binaries in globular clusters is contaminated by the cluster acceleration, but kinematic effects can be removed given knowledge of the PSR intrinsic spin down.
Asserting zero spin-down for the GC BH+PSR candidate \meerkat, the smallest physical value, gives a robust upper-bound $\kappa \leqslant 3.95$ at 95\% confidence.
Estimating this PSR spin down from a calibrator ``gold sample'' of Galactic field PSRs matched in spin period gives $\kappa = 0.82 \pm 1.67$ and $\kappa = 0.75 \pm 1.54$ depending on adopted $H_0$.
A definitive measurement of $\kappa > 0$ would provide striking evidence of the ``scale breaking'' possible within non-linear theories like GR, contributing deep insight into the astrophysical nature of BHs.
Our results show that $\kappa=3$ for \meerkat~will be excluded or detected in $95\%$ of measurements at the 3$\sigma$ confidence level by 2031 (2036 for $5\sigma$).
Near-future experiments such as the Square Kilometer Array will further strengthen the calibrator sample and are expected to observe $\sim 2$ BH+PSR binaries with pulse periods of milliseconds. 

\begin{acknowledgments}
KC thanks David C. and Marzia C. Schainker for their hospitality during the preparation of this work and K.~Schmidt (ASU) for discussion.
GBB acknowledges support from JWST-AR-04695 and JWST-AR-10799, which were provided through a grant from the STScI under
NASA contract NAS5-03127.
CC acknowledges support
from the Kavli Institute for Cosmology at Cambridge and the Beus Center for Cosmic Foundations at Arizona State University. 
Anthropic's Claude (Opus 5.5) was used to locate pertinent references, debate analysis approach, verify longhand calculations, verify catalog pulls, and review drafts of this manuscript.
All manuscript text, derivations, analysis code, and figures were produced manually by the authors.
RAW acknowledges support from NASA JWST Interdisciplinary Scientist grants NAG5-12460, NNX14AN10G and 80NSSC18K0200 from GSFC.
JWR acknowledges funding from U.S. Department of Energy Grant No. DE-SC0016021.
NF acknowledges funding from the U.S. Department of Energy Grant No. DE-SC0010008.
L.T. acknowledges support from JSPS Kakenhi grants 24K22878, 26K17136, JST PRESTO JPMJPR26287514, and Royal Society International Collaboration Award Round 2 (Japan) 252140.
This material is based upon work supported by the National Science Foundation Graduate Research Fellowship Program under Grant Nos. 2139433 and 2233001.
Any opinions, findings, and conclusions or recommendations expressed in this material are those of the authors and do not necessarily reflect the views of the National Science Foundation.
\end{acknowledgments}

\appendix
\section{CMB Dependent Quantities}
\label{sec:cmb}
The coordinates and motion of \meerkat~were assumed to be that of NGC~1851. The position $\left( \alpha, \delta \right) = \left( \rm 05^h\,14^m\,06.76^s, -40^\circ\,02'\,47.6'' \right)$ and radial velocity $v_{r} = \rm 320.5~km\,s^{-1}$ from \citet{Harris1996, Harris2010} were used along with the distance $d = \rm 11.66~kpc$ and proper motion $\left(\mu_{\alpha*}, \mu_\delta \right) = \left( 2.128, -0.646 \right)\,\rm mas~yr^{-1}$ from \citet{Libralato2022}.
The velocity of NGC~1851 with respect to the solar system barycenter (SSB) is, as IRCS cartesian vector, $\mathbf{v}_{\rm NGC,SSB} = \left( -71.05, 241.33, -233.55 \right)~\rm km~s^{-1}$.
To account for the motion of the SSB relative to the CMB, the CMB dipole moment $\left| \mathbf{v}_{\rm SSB,CMB} \right| = 369.82~\rm km~s^{-1}$ from \citet{PlanckDipole} was used, which points towards the Galactic coordinates $(l,b) = (264.021^\circ, 48.253^\circ)$.
As a cartesian ICRS vector this is
\begin{align}
  \mathbf{v}_{\rm SSB,CMB} = \left( -359.01, 76.69, -44.71 \right)~\rm km~s^{-1}.
\end{align}
The velocity of NGC~1851 with respect to the CMB was then computed as
\begin{align}
  \vec{V}_{\rm CMB} &= \mathbf{v}_{\rm NGC,SSB} + \mathbf{v}_{\rm SSB/CMB},
\end{align}
which becomes $\left( -430.05, 318.02, -278.26 \right)~\rm km~s^{-1}$.
Its magnitude is
\begin{align}
  V_{\rm CMB} = 602.92~\rm km~s^{-1}, \label{eqn:mag_V_CMB}
\end{align}
and its line-of-sight component $\vec{V}_{\rm CMB}\cdot\hat{n}$ is
\begin{align}
  v_{\rm CMB} = 352.14~\rm km~s^{-1}. \label{eqn:v_CMB}
\end{align}

\section{Theory}
\label{sec:theory}
The solution of \citet{CrokerCohon26} (CC26) demonstrates cosmological HMS mass evolution is permissible within GR.
The HMS mass impacts geodesic motion of test particles.
In order that the Equivalence Principle remain satisfied, the inertial mass of a CCBH must also have temporal evolution matched to the HMS mass.
For numerical bounds in this section, we adopt the largest theoretically motivated $\kappa = 3$.

Let $m_{\rm p}$ and $m_{\rm c}$ be the masses of the pulsar and companion, with velocities $\vec{v}_{\rm p}$ and $\vec{v}_{\rm c}$, respectively.
We begin with the $L_{\rm PN}$ Lagrangian in \citet[][Eqns. 2.1a-c]{DamourDeruelle85},
\begin{align}
  L_{\rm PN} := \frac{m_{\rm p}\vec{v}_{\rm p}^2}{2} + \frac{m_{\rm c}\vec{v}_{\rm c}^2}{2} + \frac{Gm_{\rm p}m_{\rm c}}{R} + \frac{L_2}{c^2}, \label{eqn:raw_lagrangian}
\end{align}
where $R$ is the magnitude of the relative separation vector $\vec{R}$ and $L_2$ encodes the 1PN corrections to the Newtonian Kepler problem.
CC26 describes a cosmologically coupled point mass at rest with respect to an isotropic CMB, receiving isotropic radial cosmological flux.
Assuming that \meerkat~is a binary PSR+BH system, we cosmologically couple \cref{eqn:raw_lagrangian} by introducing cosmological scale-factor dependence to the companion mass $m_{\rm c} \propto a^\kappa$ such that
\begin{align}
  \dot{m}_{\rm c} &= \kappa Hm_{\rm c},
\end{align}
where $H :=\dot{a}/a$ is the Hubble rate and $\kappa$ is the constant cosmological coupling strength.
The absence of explicit momentum transfer terms in \cref{eqn:raw_lagrangian} then implies that the harmonic coordinate system used to write \cref{eqn:raw_lagrangian} supplies the (non-orthonormal) coordinate basis for the $\vec{v}_i$ that is at rest with respect to the CMB.
Because CC26 establishes that $\kappa_{\rm PSR} \leqslant 10^{-3}$, we continue to regard $m_{\rm p}$ as constant.

\subsection{Cosmological Coupling as Hamiltonian Perturbation}
Our goal is to import as many \citet{DamourDeruelle85} results as possible.
We will achieve this by encoding the impact of cosmological coupling, to the required accuracy, as a perturbation $\Delta\mathcal{H}$ to the center-of-momentum (CM) frame 1PN Hamiltonian $\mathcal{H}_{1PN}$.
We will then compute the altered temporal dynamics via Poisson brackets against this perturbed Hamiltonian, accounting for explicit time-dependence.

We begin with the characterization of orbital motion in terms of binary relative separation 
\begin{align}
  \vec{R} := \vec{r}_{\rm p} - \vec{r}_{\rm c},
\end{align}
relative velocity $\vec{V} := \dot{\vec{R}}$, and center of rest mass position (CRM, which is not the 1PN center of total energy position),
\begin{align}
  \vec{X} := \frac{m_{\rm p}\vec{r}_{\rm p} + m_{\rm c}\vec{r}_{\rm c}}{M}. \label{eqn:defn_CM}
\end{align}
In these variables, but with time-dependent $m_{\rm c}$, the 1PN Lagrangian becomes 
\begin{align}
  L_{\rm PN} = \bigg[ \frac{M}{2}\dot{\vec{X}}^2 + &\frac{\mu}{2}\dot{\vec{R}}^2 + \frac{G\mu M}{R} + \frac{L_2}{c^2}\bigg] \notag \\
  &+  \frac{(m_{\rm p}\dot{m}_{\rm c}R)^2}{2M^3} + \dot{\vec{X}}\cdot\vec{R}\frac{m_{\rm p}\dot{m}_{\rm c}}{M} , \label{eqn:L_PN}
\end{align}
where $L_2[\vec{R}, \vec{\dot{X}}, \vec{\dot{R}}, m_{\rm c}(t)]$ contains the 1PN corrections.
Note that $\dot{\vec{X}}$ is the binary's motion relative to the CMB, because we have not changed frames, only switched to relative variables.
The terms in square braces give a Poincar\'e invariant model if mass is fixed.
The terms with $\dot{m}_{\rm c}$ are not invariant.
Let dots denote contributions from $L_2$.
The conjugate momenta for \cref{eqn:L_PN} are,
\begin{align}
  \pi_{\vec R} &= \mu \dot{\vec{R}} + \dots \\ 
  \pi_{\vec X} &= M\dot{\vec{X}} + \frac{m_{\rm p}\dot{m}_{\rm c}}{M}\vec{R} + \dots, \label{eqn:X_momentum} 
\end{align}
with $\pi_\vec{X}$ conserved.
The Hamiltonian follows from Legendre transformation, 
\begin{align}
  \mathcal{H} = \left[\frac{\pi_{\vec{R}}^2}{2\mu} - \frac{GM\mu}{R} - \frac{L_2}{c^2}\right] + \frac{\pi_{\vec{X}}^2}{2M} - \frac{m_{\rm p}\dot{m}_{\rm c}}{M^2} \vec{R}\cdot\vec{\pi}_\vec{X}.
\end{align}
The portion of the Hamiltonian in square braces is identical in form to that obtained from the Poincar\'e invariant portion of \cref{eqn:L_PN} at fixed mass, after transforming into the binary center of momentum frame. 
In addition, the term containing $\vec{R}\cdot\pi_\vec{X}$ produces $m_{\rm p}\dot{m}_{\rm c}(\vec{R}\cdot\vec{\dot{\vec{X}}})/M$ plus negligible $\mathcal{O}(H^2)$ corrections.
To the required accuracy then, the center of momentum binary problem as considered by \citet{DamourDeruelle85} can be cosmologically coupled by the addition of a perturbing potential,%
\footnote{We drop a fractional correction $(V_{\rm CMB}/c)^2\sim 10^{-6}$ acquired from the Lorentz $\gamma$ in moving $H$ from the CMB rest frame into the binary center of momentum frame.
An additional correction appears in the separation vector.
Let $t'$ be the binary harmonic time coordinate, and $t$ be CMB rest frame time, and work in 1+1 for clarity.
Then $r_i(t') = r_i(t)[1 + v_i V_{\rm CMB}/c^2 + \cdots] + V_{\rm CMB}t$ from the Lorentz transformation.
The difference $R$ acquires a fractional correction $\sim V V_{\rm CMB}/c^2 \sim \epsilon V_{\rm CMB}/V \sim 10^{-6}$, where $\epsilon$ is defined in \cref{eqn:epsilon}.
We also drop this term.}
\begin{align}
  \Delta\mathcal{H} := -\kappa H\mu \left(\vec{R}\cdot\vec{V}_{\rm CMB}\right), \label{eqn:Delta_H}
\end{align}
where $\vec{V}_{\rm CMB}$ is the binary center-of-momentum velocity relative to the CMB.
The full 1PN Hamiltonian with variable-mass then has the same form as the Poincar\'e invariant 1PN Hamiltonian, plus a perturbation, 
\begin{align}
  \mathcal{H} = \mathcal{H}_{\rm 1PN} + \Delta\mathcal{H}.
\end{align}
This allows us to use expressions for the 1PN conserved energy and angular momentum per unit reduced mass, $E$ and $\vec{J} := \vec{L}/\mu$ respectively, with the variable mass substituted.
From this point forward, overset dots will denote derivatives with respect to binary center of momentum time (not CMB time).

Before we can make predictions for data, we need to characterize additional kinematic contributions to $\dot{P}_{\rm b,obs}$, compute the additional intrinsic orbital period change $\dot{P}_{\rm b,int}$ due to CC, and determine how other observables used to infer the binary's masses, like the rate of periastron advance $\dot{\omega}$ and Einstein delay $\gamma_{\rm E}$, evolve when $\dot{m}_{\rm c} \neq 0$.

\subsection{Kinematic acceleration}
\label{sec:CC_kinematic}
Because $\vec{V}_{\rm CMB}$ is not constant, but \cref{eqn:X_momentum} is conserved in the center-of-momentum frame,
\begin{align}
  \vec{\dot{V}}_{\rm CMB} = -\frac{\kappa H}{1+q}\vec{V}_{\rm CMB} - \frac{\d}{\d t}\left[\frac{m_{\rm p}\dot{m}_{\rm c}}{M}\vec{R} + \dots\right],
\end{align}
The term in square braces contributes $\mathcal{O}(H^2)$ negligible corrections and a term $\propto \vec{V}$ that orbit-averages to zero at the same precision.
Thus, \cref{eqn:period_variation} acquires an additional secular term, 
\begin{align}
  \frac{a_{\rm CMB}}{c} = -\frac{\kappa H v_{\rm CMB}}{c(1+q)}, \label{eqn:a_CMB}
\end{align}
where $v_{\rm CMB}$ is the projection of $\vec{V}_{\rm CMB}$ along the line-of-sight given by \cref{eqn:v_CMB}.
The term contributes $\sim 10^{-21}\,\rm s^{-1}$.
Note that $a_{\rm gal}$ is the relative acceleration of the SSB and pulsar CM, due to their positions within the Galactic gravitational potential.
It cannot absorb \cref{eqn:a_CMB}, which would be present even in the absence of the galaxy.

\subsection{Orbital period change}
\label{sec:CC_period_change}
To compute the contribution to $\dot{P}_{\rm b}^{\rm int}/P_{\rm b}$ from CC, we start from the 1PN generalization of Kepler's 3rd law given by \citet[][Eqs. 2.13, 3.6d]{DamourDeruelle85},
\begin{align}
  \frac{2\pi}{P_{\rm b}} &= \frac{(-2E)^{3/2}}{GM}\left[1 - \frac{E}{4c^2}\left(\nu - 15\right)\right] \label{eqn:kepler3}
\end{align}
and the 1PN conserved energy at fixed mass in the center of momentum frame,
\begin{align}
  \begin{split}
    E = &\frac{V^2}{2} - \frac{GM}{R} + \frac{3(1-3\nu)}{8c^2}V^4\\
    & + \frac{GM}{2Rc^2}\left[(3+\nu)V^2 + \nu\left(\frac{\vec{V}\cdot\vec{R}}{R}\right)^2 + \frac{GM}{R}\right].   
  \label{eqn:E}
  \end{split}
\end{align}
As we proceed, it will become apparent that the $L_2$ corrections need not be explicitly expanded.
To track the order of terms appearing in subsequent calculations, we introduce the usual small dimensionless parameter,
\begin{align}
  \epsilon := \frac{GM}{Rc^2} \sim \left(\frac{V}{c}\right)^2. \label{eqn:epsilon}
\end{align}
Taking a logarithmic derivative of \cref{eqn:kepler3} with respect to time and averaging over the 0PN orbit $[t', t'+P_{\rm b}]$, following \citet{Peters64} gives,
\begin{align}
  \left<\frac{\dot{P}_{\rm b}}{P_{\rm b}}\right> = \left[-\frac{3}{2}\left<\frac{\dot{E}}{E}\right> + \left<\frac{\dot{M}}{M}\right>\right]\left[1 + \mathcal{O}(\epsilon)\right], \label{eqn:period_decay}
\end{align}
where we have discarded terms $\mathcal{O}(\epsilon^2)$.
Evaluation of the total mass (and reduced mass) is straightforward,
\begin{align}
  \frac{\dot{M}}{M} = \frac{\kappa H}{1+q} \qquad
  \frac{\dot{\mu}}{\mu} = \kappa H \left(1 - \frac{1}{1+q}\right). \label{eqn:logarithmic_mass_shifts}
\end{align}
To get $\dot{E}$, we use the Poisson bracket \citep[e.g.][Eqn.~9-94]{Goldstein2nd},
\begin{align}
  \dot{E} = \{E, \Delta\mathcal{H}\} + \frac{\partial E}{\partial t}\bigg|_{\pi_{\vec{R}}, \vec{R}}, \label{eqn:bracket}
\end{align}
leveraging linearity of the bracket and that $\{E, \mathcal{H}_{\rm 1PN}\} = 0$.
Next express $E$ in canonical variables,
\begin{align}
  E = \frac{\pi_{\vec{R}}^2}{2\mu^2} - \frac{GM}{R} + \mathcal{O}(\epsilon V^2),
\end{align}
to enable the evaluation of \cref{eqn:bracket},
\begin{align}
  \dot{E} = \frac{\kappa H(\pi_{\vec{R}}\cdot\vec{V}_{\rm CMB})}{\mu} -\frac{\dot{\mu}\pi_{\vec{R}}^2}{\mu^3} - \frac{G\dot{M}}{R} + \mathcal{O}(\kappa H\epsilon V^2).
\end{align}
Although $\vec{V}_{\rm CMB}$ is not constant, its derivative introduces $H$, so its time-evolution introduces negligible $\mathcal{O}(H^2)$ terms.

In general, to extract the slow Hubble expansion (via time-dependent mass) from orbit-averages,
\begin{align}
  \left<\dot{E}\right> = \frac{1}{P_{\rm b}}\int_{t'}^{t' + P_{\rm b}} \frac{\kappa H(\pi_{\vec{R}}\cdot\vec{V}_{\rm CMB})}{\mu}\,\d t'' + \cdots
\end{align}
Taylor expand $H$ and mass terms like $\mu$ about $t'$,
\begin{align}
  H(t'') = H(t') + \dot{H}(t')(t'' - t') + \cdots,
\end{align}
and use Friedmann's equations to replace $\dot{H} \sim H^2$ and note $t' - t'' \leqslant P_{\rm b}$ always,
\begin{align}
  \frac{\kappa H\left[1 + \mathcal{O}(HP_{\rm b})\right]}{\mu P_{\rm b}}\int_{t'}^{t' + P_{\rm b}}(\pi_{\vec{R}}\cdot\vec{V}_{\rm CMB})\,\d t'' + \cdots,
\end{align}
expand similarly for $\mu$, and drop $\mathcal{O}(H^2)$.
Although the required average is of $E$'s logarithmic derivative, because $E$ is constant along the 0PN orbit, it commutes through the average of $\dot{E}/E$.
The $\vec{V}_{\rm CMB}$ term then becomes a total derivative that vanishes upon orbit-average.
Substituting \cref{eqn:logarithmic_mass_shifts} gives,
\begin{align}
  \left<\frac{\dot{E}}{E}\right> = -\frac{\kappa H}{E(1+q)} \left[q\left<V^2\right> + \left<\frac{GM}{R}\right>\right] + \mathcal{O}(\kappa H\epsilon) \label{eqn:dotE_over_E}. 
\end{align}
The Virial Theorem applied to the 0PN orbit gives
\begin{align}
  \left<\frac{GM}{R}\right> = \left<V^2\right> \qquad E = -\frac{1}{2}\left<V^2\right> \qquad \left<\frac{\dot{E}}{E}\right> = 2\kappa H.\label{eqn:virial}
\end{align}
Substituting \cref{eqn:logarithmic_mass_shifts,eqn:virial,eqn:dotE_over_E} into \cref{eqn:period_decay} finally gives
\begin{align}
  \left<\frac{\dot{P}_{\rm b}^{\rm CC}}{P_{\rm b}}\right> = -\kappa H\left[\left(3 - \frac{1}{1+q}\right) + \mathcal{O}(\epsilon)\right]. \label{eqn:decay_cons}
\end{align}
This is the Keplerian result with adiabatic invariants found by \citet{CrokerNishimura20}, generalized to a single cosmologically coupled mass and with an explicit error bound.
Using the projected orbit semi-major axis reported by \citet{meerkat24} to establish a conservative bound, the relative 1PN correction for \meerkat~is,
\begin{align}
  \epsilon \leqslant 6.8 \times 10^{-7},
\end{align}
which is beyond the current timing precision of $\dot{P}_{\rm b,obs}$ by $10^4$.
We note that the $a_{\rm CMB}/c$ term is $\sim 0.1\%$ of this leading order contribution.

The effect is always present, regardless of the value of $q$.
Note that $q > 1$ is possible because, in the absence of horizons, the TOV upper-limit mass for NS stability does not set a lower-bound stellar collapse BH mass \citep[e.g.][Fig. 14]{Amendola24}.

\subsection{Periastron precession}
\label{sec:CC_periastron}
Knowledge of the mass ratio $q$ is required for accurate inference of $\kappa$ from $\dot{P}_{\rm b}$ measurements.
Measurement of the total mass is determined from periastron precession $\dot{\omega}$.
Canonical angular momentum is defined
\begin{align}
  \vec{L} := \vec{R} \times \pi_\vec{R},
\end{align}
The Poisson bracket gives,
\begin{align}
  \dot{\vec{L}} = \left\{\vec{L}, \Delta\mathcal{H}\right\} = \kappa H \mu (\vec{R}\times\vec{V}_{\rm CMB}) 
  \label{eqn:dot_vec_L}
\end{align}
and so the orbital plane receives a small torque.
To verify that the masses reported by \citet{meerkat24} are not biased by CC, we obtain the periastron precession via the Laplace-Runge-Lenz vector,
\begin{align}
  \vec{A} := \pi_{\vec{R}} \times \vec{L} - \frac{GM\mu^2}{R}\vec{R}. \label{eqn:LRL}
\end{align}
Evaluating the Poisson bracket and adding the temporal partial gives, 
\begin{align}
  \Delta\dot{\vec{A}} = \kappa H\mu\bigg[&\vec{V}_{\rm CMB}\times\vec{L} - (\vec{R}\cdot\pi_\vec{R})\vec{V}_{\rm CMB}  \notag \\
    &+ (\pi_\vec{R}\cdot\vec{V}_{\rm CMB})\vec{R}\bigg] - \frac{\vec{R}}{R}G\mu^2\dot{m}_{\rm c}(1+2q), \label{eqn:to_eval}
\end{align}
where the $\Delta$ notation encodes the deviation from the 1PN fixed-mass behavior.
Within the square braces, the first term is constant under 0PN orbit average.
The second term is the total derivative $\d R^2/\d t$ and vanishes under orbit average.
The third term is most easily averaged in index notation,
\begin{align}
  \left<(\pi_\vec{R}\cdot\vec{V}_{\rm CMB})\vec{R}\right> = \frac{\mu}{P_{\rm b}}\int_{t'}^{t' + P_{\rm b}} V^iR^j\,\d t'' V^i_{\rm CMB}.
\end{align}
Write the integrand as,
\begin{align}
  V^iR^j = \frac{1}{2}\left[(V^iR^j + V^jR^i) + (V^iR^j - V^jR^i)\right].
\end{align}
The symmetric portion is again $\d R^2/\d t$ and vanishes from the average.
The antisymmetric portion can be recognized as $\epsilon_{jik}L^k/\mu$, and so
\begin{align}
  \left<(\pi_\vec{R}\cdot\vec{V}_{\rm CMB})\vec{R}\right> = \frac{1}{2} \vec{V}_{\rm CMB}\times \vec{L}.
\end{align}
The required average in the fourth term of \cref{eqn:to_eval} is,
\begin{align}
  \frac{1}{P_{\rm b}}\int_{t'}^{t' + P_{\rm b}} \frac{\vec{R}}{R} \,\d t''.
\end{align}
\citet[][Eqns. 3.4, 4.6]{DamourDeruelle85} give the relative separation and Jacobian between binary time and eccentric anomaly $u$,
\begin{align}
  R &= a_{\rm R} (1-e\cos u) \\
  \d t'' &= \frac{P_{\rm b}}{2\pi}(1 - e\cos u)\,\d u, \label{eqn:eccentric_anomaly_jacobian}
\end{align}
where we have set the temporal and spatial eccentricities of the 1PN problem equal to the Keplerian eccentricity $e$ along the 0PN orbit.
The fourth term average becomes,
\begin{align}
  \frac{1}{2\pi a_{\rm R}}\int_0^{2\pi} \vec{R} \,\d u,
\end{align}
where we have leveraged periodicity of $\vec{R}$ to label the starting angle as 0.
Let $\hat{\vec{e}}$ point from attractive focus to periastron.
Then $\{\hat{\vec{e}}, \hat{\vec{L}} \times \hat{\vec{e}}\}$ is the orbital plane basis%
\footnote{the direction of periastron is defined on page 15, \emph{op. cit.}}
in which the components of $\vec{R}$ are given by 
\citet[][Eqs. 2.41, 2.42]{Klioner16},
\begin{align}
  \vec{R} = a_{\rm R}(\cos u - e)\hat{\vec{e}} + (a_{\rm R}\sqrt{1-e^2}\sin u)\hat{\vec{L}} \times \hat{\vec{e}}. \label{eqn:vec_R}
\end{align}
At 0PN, the basis remains fixed and performing the integration gives $-e\hat{\vec{e}}$.
Assembling all the pieces gives,
\begin{align}
  \left<\Delta\dot{\vec{A}}\right> = \kappa H\mu\left[\frac{3}{2}\left(\vec{V}_{\rm CMB} \times \vec{L}\right) + G\mu m_{\rm c}(1+2q)\vec{e}\right]. \label{eqn:dot_vec_A}
\end{align}
A component of $\dot\Delta\vec{A}$ along $\vec{e}$ indicates orbital stretch, because $\vec{A}=GM\mu^2\vec{e}$.
The magnitude of the term driving CC precession is the overlap of the other term with $\hat{\vec{e}}\times\hat{\vec{L}}$,
\begin{align}
  \frac{3\kappa H \mu}{2}(\vec{\hat{e}}\cdot\vec{V}_{\rm CMB})L \leqslant \frac{3}{2}\kappa H \mu^2\frac{V_{\rm CMB}}{V_{\rm ap}}Rc^2\epsilon.
\end{align}
Here $V_{\rm ap}$ is the relative velocity at apastron, the orbital minimum $V$.
Now, note that the time-derivative of \cref{eqn:LRL} can be written as,
\begin{align}
  \dot{\vec{A}} = \dot{\hat{\vec{e}}} \left(GM\mu^2e\right) + \hat{\vec{e}}\frac{\d}{\d t}\left(GM\mu^2e\right),
\end{align}
and we find
\begin{align}
  |\dot{\widehat{\vec{e}}}|_{\rm CC} \leqslant \frac{3V_{\rm CMB}}{2V_{\rm ap}e}\kappa H \sim 2\times 10^{-7}\,\rm deg\,yr^{-1}.
\end{align}
The value is roughly $100\times$ smaller than the $1\sigma$ errors reported by \citet{meerkat24}.
We conclude%
\footnote{We note a negligible fractional drift in $\dot{\omega}_{1PN} \propto \kappa H T_\star = 3.5\times 10^{-9}$}
that the extracted total mass $M$ is robust.

\subsection{Einstein Delay}
\label{sec:CC_einstein_delay}
The Einstein delay $\gamma_{\rm E}$ gives the magnitude of the oscillation in pulsar clock rate relative to the binary CM clock rate \citep[][Eq.(20)]{DamourDeruelle86},
\begin{align}
  \gamma_{\rm E} = \frac{e P_{\rm b}}{2\pi}\frac{G}{a_{\rm R}c^2} \frac{m_{\rm c}\left(m_{\rm p} + 2m_{\rm c}\right)}{M}.
\end{align}
Because the masses enter asymmetrically, measurement of $\gamma_{\rm E}$ can be used to determine the component masses when combined with the  precession of the periastron $\dot{\omega}$.
Take the logarithmic derivative, substitute \cref{eqn:logarithmic_mass_shifts,eqn:virial}, and 0PN orbit-average,
\begin{align}
  \left<\frac{\dot{\gamma}_{\rm E}}{\gamma_{\rm E}}\right> = \left<\frac{\dot{e}}{e}\right> + \frac{\kappa H q}{(q+1)(q+2)}. \label{eqn:dot_log_gamma}
\end{align}
To resolve the eccentricity time evolution, take a logarithmic derivative of 0PN $e^2 - 1$, and again substitute \cref{eqn:logarithmic_mass_shifts,eqn:virial},
\begin{align}
  \left<\frac{\dot{e}}{e}\right> = \frac{e^2 - 1}{L^2e^2} \vec{L}\cdot\dot{\left<\vec{L}\right>}, \label{eqn:dot_log_e}
\end{align}
where we have extracted $\vec{L}$ from the average because it is constant at 0PN.
Integration of \cref{eqn:vec_R} within \cref{eqn:dot_vec_L} over eccentric anomaly $u$ with \cref{eqn:eccentric_anomaly_jacobian} gives,
\begin{align}
  \dot{\left<\vec{L}\right>} = -\frac{3}{2}ea_{\rm R}\kappa H \mu \hat{\vec{e}}\times\vec{V}_{\rm CMB}. \label{eqn:average_dot_vec_L}
\end{align}
Assembling \cref{eqn:dot_log_gamma,eqn:dot_log_e,eqn:average_dot_vec_L} and compressing via standard Kepler relations gives
\begin{align}
  \left<\frac{\dot{\gamma}_{\rm E}}{\gamma_{\rm E}}\right> = \frac{3}{2}\frac{\kappa H (\hat{\vec{e}}\times\vec{V}_{\rm CMB}\cdot\vec{L})}{Gm_{\rm p}m_{\rm c}e} + \frac{\kappa H q}{(q+1)(q+2)}.
\end{align}
This can be put into more transparent form via the semi-latus rectum and semi-major axis relations given by \citet[][Eqs. 15.4,15.6]{LandauMechanics} and Kepler's 3rd law,
\begin{align}
  \left<\frac{\dot{\gamma}_{\rm E}}{\gamma_{\rm E}}\right> = \frac{3\kappa H \sqrt{1-e^2}}{4\pi e} \frac{(\hat{\vec{e}}\times\vec{V}_{\rm CMB}\cdot\hat{\vec{L}})}{a_{\rm R}P_{\rm b}^{-1}} + \frac{\kappa H q}{q^2+3q+2}. \label{eqn:gamma_E_drift} 
\end{align}
Using \cref{eqn:virial} and \citet[][Eq. 3.6a]{DamourDeruelle85}, $2\pi a_{\rm R}/P_{\rm b} = V_{\rm rms}$, where
\begin{align}
  V_{\rm rms} = \sqrt{2|E|} = \sqrt{2\left(\frac{GM\pi}{P_{\rm b}\sqrt{2}}\right)^{2/3}} = 171.4\,\rm km/s.
\end{align}
So the relative shift due to CC is manifestly $\mathcal{O}(\kappa H)$ for \meerkat.
We conclude that the component masses extracted with $\gamma_{\rm E}$ are robust.

\subsection{Orbital inclination and Shapiro delay}
\label{sec:CC_orbital_i}
The orbital plane is torqued because \cref{eqn:dot_vec_L} is non-zero in general.
This will change the orbital inclination angle $i$ to the line of sight $\hat{n}$.
\citet{meerkat24} use the sine of $i$ to develop bounds on $m_{\rm p}$, and trigonometric functions of $i$ enter computation of the orthometric Shapiro delay parameter $h_3$, so we bound its evolution as well.
Recall that $\vec{L}$ is orthogonal to the orbital plane, and that $\cos i = \vec{L} \cdot \hat{n}/L$.
Regard $\hat{n}$ as constant to determine the leading order CC impact and find,
\begin{align}
  \frac{\d \sin i}{\d t} = -\cot i \left[\frac{\dot{\vec{L}}\cdot\hat{n}}{L} - \frac{\vec{L}\cdot\hat{n}}{L^3}\dot{\vec{L}}\cdot\vec{L}\right].
\end{align}
Averaging over the 0PN orbit, substituting \cref{eqn:average_dot_vec_L}, and expressing in terms of $V_{\rm rms}$ gives,
\begin{align}
  \left<\frac{\d\sin i}{\d t}\right> = \frac{3e\kappa H \cot i }{2\sqrt{1-e^2} } \left(\hat{\vec{e}}\times\frac{\vec{V}_{\rm CMB}}{V_{\rm rms}}\right)\cdot\left[\hat{n} - (\hat{\vec{L}}\cdot\hat{n})\hat{\vec{L}}\right].
\end{align}
The accumulated angle over a 25 year baseline at $\kappa = 3$ is at most $3\times 10^{-8}\,\rm rad$.
The reported $i = 52^{+6}_{-5}\,\rm deg$ which, at $1\sigma$, is already $\sim 0.1\,\rm rad$ uncertain.
We conclude that inferences with orbital inclination angle and Shapiro delay are unaffected. 

\subsection{Extension of CC to higher PN orders}
\label{sec:higher_PN}
While we have worked only to 1PN order explicitly in this appendix because 1PN effects are currently what is measured and constrained for \meerkat, the Poisson bracket formalism immediately generalizes to higher PN orders.
The essential point is that the frame-dependence of CC is encoded via the Hamiltonian perturbation \cref{eqn:Delta_H}, whereas the impact of explicitly time-dependent companion mass $m_{\rm c}$ enters only through a temporal partial at fixed phase-space variables.

Explicitly, let $Y$ be some generic phase-space function with explicit time-dependence.
Evolved along the Hamiltonian flow, it is always true that,
\begin{align}
  \dot{Y} = \left\{Y, \mathcal{H}\right\} + \frac{\partial Y}{\partial t}\bigg|_{Q_i,\pi_i}.
\end{align}
Suppose the Hamiltonian is given through 2PN, i.e. it includes the textbook 1PN, 1.5PN, and 2PN pieces.
Then the conserved $E$ through 2PN continues to commute with $\mathcal{H}_{\rm 2PN}$ and so,
\begin{align}
  \dot{E}_{\rm 2PN} = \left\{E_{\rm 2PN}, \Delta\mathcal{H}\right\} + \frac{\partial E_{\rm 2PN}}{\partial t}\bigg|_{Q_i,\pi_i}. 
\end{align}
The conserved $\vec{J}$ through 2PN also continues to commute with $\mathcal{H}_{\rm 2PN}$ and so,
\begin{align}
  \dot{\vec{J}}_{\rm 2PN} = \left\{\vec{J}_{\rm 2PN}, \Delta\mathcal{H}\right\} + \frac{\partial \vec{J}_{\rm 2PN}}{\partial t}\bigg|_{Q_i,\pi_i},
\end{align}
provided that CC does not alter the spin of the coupled object.
Regardless of the PN order being interrogated, the surviving Poisson bracket always contributes frame-dependence corrections bounded above by $\mathcal{O}(\kappa H V_{\rm CMB}/V_{\rm rms})$ relative to the decoupled effect, as we have just seen explicitly at 1PN.
Meanwhile, the temporal partial contributions always introduce fractional corrections from explicit temporal mass evolution at $\mathcal{O}(\kappa H)$, regardless of PN order.

\subsection{Independence from Radiative Losses}
Gravitational radiation in GR results from the time-varying quadrupole moment and enters at 2.5PN.
In particular, for the decoupled problem, with $\mathcal{H}_{\rm 2.5PN}$ taken from \citet[][Eqs. 6.108, 6.109]{SchaferJaranowski24},
\begin{align}
  \left<\dot{P_{\rm b}}\right> = \left<\left\{P_{\rm b}, \mathcal{H}_{\rm 2.5PN}\right\}\right> + \frac{\partial P_{\rm b}}{\partial t}\bigg|_{Q_i,\pi_i},
\end{align}
gives the usual \citet{Peters64} radiative decay at fixed mass when averaged over the 0PN orbit,
\begin{align}
    \left<\dot{P}_{\rm b}^\mathrm{GW}\right> &= -\frac{192\pi}{5}\left(\frac{G}{c^2}\right)^{5/2}\frac{\mu(m_{\rm c} + m_{\rm p})^{3/2}}{a_{\rm R}^{5/2}}F(e) \label{eqn:Peters_OG} \\
    F(e) &:= \frac{1}{(1-e^2)^{7/2}}\left(1 + \frac{73}{24}e^2 + \frac{37}{96}e^4\right),
\end{align} 
because $\partial P_{\rm b}/\partial t = 0$ at fixed phase-space in the fixed-mass problem.
As we have shown in \ref{sec:CC_period_change}, the CC orbital decay is nothing but the additional pieces to this bracket,
\begin{align}
  \dot{P}_{\rm b}^{\rm CC} &= \left<\left\{P_{\rm b}, \Delta\mathcal{H}\right\} + \frac{\partial P_{\rm b}}{\partial t}\bigg|_{Q_i,\pi_i}\right>,
\end{align}
which happen to be non-vanishing at much lower order.
Of course, within this bracket, at 2.5PN $P_{\rm b}$ acquires fractional corrections through $\mathcal{O}(\epsilon^{2.5})$.
But the relative pattern of suppression for CC impacts remains unchanged.
This is consistent with order-of-magnitude expectations from existing treatments of time-varying masses in the CM frame by \citet[][]{HolgadoRicker19}, when the variation time-scale is extremely slow compared to the orbit period.

For computational convenience, \cref{eqn:Peters_OG} can be expressed in terms of $P_{\rm b}$ using \cref{eqn:kepler3},
\begin{align}
    \dot{P}_{\rm b}^\mathrm{GW} &= -\frac{192\pi}{5}\left(\frac{2\pi G \mathcal{M}}{P_{\rm b} c^3}\right)^{5/3} F(e),  \label{eqn:peters_dotPb}
\end{align} 
where $\mathcal{M} := (m_{\rm c}m_{\rm p})^{3/5}/(m_{\rm c} + m_{\rm p})^{1/5}$ is the chirp mass.

\section{Forecasting}
\label{sec:forecast}
Model any realization of $\kappa$ measurement after a baseline $T$ via a normal distribution with variance $\sigma_T$ that depends on $T$,
\begin{align}
  K(\kappa, T) := \frac{1}{\sigma_T \sqrt{2\pi}} \exp\left[-\frac{(\kappa - \kappa_{\rm true})^2}{2\sigma_T^2}\right].
\end{align}
Then, the probability of measuring any $\kappa \geqslant \kappa_0$ after baseline $T$ is
\begin{align}
  p(\kappa \geqslant \kappa_0) = \int_{\kappa_0}^\infty K(\kappa, T)\,\d \kappa. \label{eqn:confident_detection}
\end{align}
Suppose that we want to claim an $N_\sigma$ or better detection of $\kappa_{\rm true}$.
Then the distance from zero of the measured $\kappa$ must be at least $N_\sigma\sigma_T$, i.e.
\begin{align}
  \kappa_0 = N_\sigma \sigma_T.
\end{align}
Perform the substitution $u := (\kappa - \kappa_{\rm true})/\sigma_T\sqrt{2}$ to bring \cref{eqn:confident_detection} into standard form,
\begin{align}
  p(\kappa \geqslant \kappa_0) &= \frac{1}{\sqrt{\pi}}\int_{\frac{N_\sigma}{\sqrt{2}} - \frac{\kappa_{\rm true}}{\sigma_T\sqrt{2}}}^\infty \exp\left(-u^2\right)\,\d u \\
  &= \frac{1}{2}\mathrm{erfc}\left(\frac{N_\sigma}{\sqrt{2}} - \frac{\kappa_{\rm true}}{\sigma_T\sqrt{2}}\right),
\end{align}
where $\mathrm{erfc(z)}$ is unity ($K$ integrated over $(-\infty, \infty)$) minus the cumulative distribution function for the standard gaussian PDF.
The 2 comes from even symmetry of the integrand.
Our goal becomes to invert this relation,
\begin{align}
  \sigma_T = \frac{\kappa_{\rm true}}{N_\sigma - \mathrm{erfc}^{-1}(2 p)\sqrt{2}}, \label{eqn:invert_me}
\end{align}
and solve for the $T$ required to achieve some given confidence $p$ on our $\kappa$ measurement,

The linear error propagation formula applied to \cref{eqn:decay_cons} (at fixed $n(q)$ for simplicity, because uncertainty in $q$ is subdominant)
\begin{align}
  \left(\frac{\Delta\dot{P}_{\rm b}^{\rm CC}}{\dot{P}_{\rm b}^{\rm CC}}\right)^2 = \left(\frac{\sigma_T}{\kappa}\right)^2 + \left(\frac{\Delta H}{H}\right)^2 + \left(\frac{\Delta P_{\rm b}}{P_{\rm b}}\right)^2.
\end{align}
By far, the leading contribution is from $\Delta\dot{P}_{\rm b}^{\rm CC}$.
Solving for $\sigma_T$ in this regime and substituting \cref{eqn:decay_cons} for $\dot{P}_{\rm b}^{\rm CC}$ gives,
\begin{align}
  \sigma_T = \frac{\Delta\dot{P}_{\rm b}^{\rm CC}}{HP_{\rm b}n(q)}. \label{eqn:sigma_T}
\end{align}
We apply linear error propagation to \cref{eqn:difference_trick} and resolve $\Delta\dot{P}_{\rm b}^{\rm CC}$ in terms of the dominant uncertainties in observables,
\begin{align}
  \left[\Delta\left(\frac{\dot{P}_{\rm b}^{\rm CC}}{P_{\rm b}}\right)\right]^2 = \left[\Delta\left(\frac{\dot{P}_{\rm b,obs}}{P_{\rm b}}\right)\right]^2 + \left[\Delta\left(\frac{\dot{P}_{\rm int}}{P}\right)\right]^2
\end{align}
Substitution into \cref{eqn:sigma_T} gives,
\begin{align}
  \sigma_T = \frac{1}{Hn(q)}\sqrt{\frac{(\Delta \dot{P}_{\rm b,obs})^2}{P_{\rm b}^2} + \frac{(\Delta\dot{P}_{\rm int})^2}{P^2}}.
\end{align}
where we have leveraged the exquisite timing precision to regard $P_{\rm b}$ and $P$ as constants that commute freely with $\Delta$.
The uncertainty in $\dot{P}_{\rm b,obs}$ is modeled analogously to the \citet[][Table II]{DamourTaylor92} analysis of PSR B1913+16,
\begin{align}
  \Delta \dot{P}_{\rm b,obs} = \lambda T^{-\alpha} \qquad (2 \leqslant \alpha \leqslant 2.5). \label{eqn:DT_resolution}
\end{align}
Here $\lambda$ and $\alpha$ are constants determined by detector performance and observation cadence.
Let $T_\star$ be the \citet{meerkat24} observation baseline.
Then, given $\alpha$, $\lambda$ is determined by the uncertainty in $\dot{P}_{\rm b,obs}$ listed in \cref{tbl:meerkat},
\begin{align}
  \lambda = \Delta\dot{P}_{\rm b,obs}^\star T_\star^\alpha. \label{eqn:lambda}
\end{align}
Equating \cref{eqn:invert_me,eqn:sigma_T}, substituting \cref{eqn:DT_resolution,eqn:lambda} and solving for relative baseline increase gives,
\begin{widetext}
  \begin{align}
    \frac{T}{T_\star} = \left\{\left(\frac{P_{\rm b}}{\Delta \dot{P}_{\rm b,obs}^\star}\right)^2\left[\left(\frac{\kappa_{\rm true}Hn(q)}{N_\sigma - \mathrm{erfc}^{-1}(2p)\sqrt{2}}\right)^2 - \frac{\Delta\dot{P}_{\rm int}^2}{P^2}\right]\right\}^{-1/(2\alpha)}. \label{eqn:forecast_baseline}
  \end{align}
\end{widetext}
Finally, we compute \cref{eqn:forecast_baseline} over our $10^6$ parameter draws, and $10^6$ distinct realizations of the $N=32$ field pulsar calibrator sample (draws with repeats) to get $10^6$ estimates of $\Delta\dot{P}_{\rm int}$.
Confidence intervals at $16\%$ and $84\%$ for the forecast dates in \cref{tbl:dakine} typically range from a few months for $\kappa = 3$ to a few years for $\kappa = 1$, with increasingly asymmetric spread upward for high-significant $\kappa = 1$ forecasts.

\bibliography{merged}{}

\begin{thebibliography}{}
\expandafter\ifx\csname natexlab\endcsname\relax\def\natexlab#1{#1}\fi
\providecommand{\url}[1]{\href{#1}{#1}}
\providecommand{\dodoi}[1]{doi:~\href{http://doi.org/#1}{\nolinkurl{#1}}}
\providecommand{\doeprint}[1]{\href{http://ascl.net/#1}{\nolinkurl{http://ascl.net/#1}}}
\providecommand{\doarXiv}[1]{\href{https://arxiv.org/abs/#1}{\nolinkurl{https://arxiv.org/abs/#1}}}

\bibitem[{G. {Agazie} {et~al.}(2023){Agazie}, {Anumarlapudi}, {Archibald},
  {Arzoumanian}, {Baker}, {B{\'e}csy}, {Blecha}, {Brazier}, {Brook},
  {Burke-Spolaor}, {Burnette}, {Case}, {Charisi}, {Chatterjee},
  {Chatziioannou}, {Cheeseboro}, {Chen}, {Cohen}, {Cordes}, {Cornish},
  {Crawford}, {Cromartie}, {Crowter}, {Cutler}, {Decesar}, {Degan}, {Demorest},
  {Deng}, {Dolch}, {Drachler}, {Ellis}, {Ferrara}, {Fiore}, {Fonseca},
  {Freedman}, {Garver-Daniels}, {Gentile}, {Gersbach}, {Glaser}, {Good},
  {G{\"u}ltekin}, {Hazboun}, {Hourihane}, {Islo}, {Jennings}, {Johnson},
  {Jones}, {Kaiser}, {Kaplan}, {Kelley}, {Kerr}, {Key}, {Klein}, {Laal}, {Lam},
  {Lamb}, {Lazio}, {Lewandowska}, {Littenberg}, {Liu}, {Lommen}, {Lorimer},
  {Luo}, {Lynch}, {Ma}, {Madison}, {Mattson}, {McEwen}, {McKee}, {McLaughlin},
  {McMann}, {Meyers}, {Meyers}, {Mingarelli}, {Mitridate}, {Natarajan}, {Ng},
  {Nice}, {Ocker}, {Olum}, {Pennucci}, {Perera}, {Petrov}, {Pol}, {Radovan},
  {Ransom}, {Ray}, {Romano}, {Sardesai}, {Schmiedekamp}, {Schmiedekamp},
  {Schmitz}, {Schult}, {Shapiro-Albert}, {Siemens}, {Simon}, {Siwek}, {Stairs},
  {Stinebring}, {Stovall}, {Sun}, {Susobhanan}, {Swiggum}, {Taylor}, {Taylor},
  {Turner}, {Unal}, {Vallisneri}, {van Haasteren}, {Vigeland}, {Wahl}, {Wang},
  {Witt}, {Young}, \& {Nanograv Collaboration}}]{NANOgravFirst}
{Agazie}, G., {Anumarlapudi}, A., {Archibald}, A.~M., {et~al.} 2023,
  \bibinfo{title}{{The NANOGrav 15 yr Data Set: Evidence for a
  Gravitational-wave Background},} \apjl, 951, L8,
  \dodoi{10.3847/2041-8213/acdac6}

\bibitem[{S.~P. {Ahlen} {et~al.}(2025){Ahlen}, {Aviles}, {Cartwright},
  {Croker}, {Elbers}, {Farrah}, {Fernandez}, {Niz}, {Rohlf}, {Tarl{\'e}},
  {Windhorst}, {et~al.}}]{Ahlen25}
{Ahlen}, S.~P., {Aviles}, A., {Cartwright}, B., {et~al.} 2025,
  \bibinfo{title}{{Positive Neutrino Masses with DESI DR2 via Matter Conversion
  to Dark Energy},} PRL, 135, 081003, \dodoi{10.1103/yb2k-kn7h}

\bibitem[{J. Alsing {et~al.}(2018)Alsing, Silva, \& Berti}]{Alsing2017}
Alsing, J., Silva, H.~O., \& Berti, E. 2018, \bibinfo{title}{{Evidence for a
  maximum mass cut-off in the neutron star mass distribution and constraints on
  the equation of state},} \mnras, 478, 1377, \dodoi{10.1093/mnras/sty1065}

\bibitem[{L. {Amendola} {et~al.}(2024){Amendola}, {Rodrigues}, {Kumar}, \&
  {Quartin}}]{Amendola24}
{Amendola}, L., {Rodrigues}, D.~C., {Kumar}, S., \& {Quartin}, M. 2024,
  \bibinfo{title}{{Constraints on cosmologically coupled black holes from
  gravitational wave observations and minimal formation mass},} \mnras,
  \dodoi{10.1093/mnras/stae143}

\bibitem[{R. {Andrae} {\&} K. {El-Badry}(2023){Andrae} \&
  {El-Badry}}]{AndraeElbadry23}
{Andrae}, R., \& {El-Badry}, K. 2023, \bibinfo{title}{{Constraints on the
  cosmological coupling of black holes from Gaia},} {\aap}, 673, L10,
  \dodoi{10.1051/0004-6361/202346350}

\bibitem[{E.~D. Barr {et~al.}(2024)Barr, Dutta, Freire, Cadelano, Gautam,
  Kramer, Pallanca, Ransom, Ridolfi, Stappers, Tauris, Krishnan, Wex, Bailes,
  Behrend, Buchner, Burgay, Chen, Champion, Chen, Corongiu, Geyer, Men,
  Padmanabh, \& Possenti}]{meerkat24}
Barr, E.~D., Dutta, A., Freire, P. C.~C., {et~al.} 2024, \bibinfo{title}{A
  pulsar in a binary with a compact object in the mass gap between neutron
  stars and black holes,} Science, 383, 275, \dodoi{10.1126/science.adg3005}

\bibitem[{C.~M. {Bender} {\&} S.~A. {Orszag}(1978){Bender} \&
  {Orszag}}]{BenderOrszag78}
{Bender}, C.~M., \& {Orszag}, S.~A. 1978, {Advanced Mathematical Methods for
  Scientists and Engineers}

\bibitem[{R.~S. Booth {et~al.}(2009)Booth, de~Blok, Jonas, \&
  Fanaroff}]{MeerKAT09}
Booth, R.~S., de~Blok, W. J.~G., Jonas, J.~L., \& Fanaroff, B. 2009,
  \bibinfo{title}{MeerKAT Key Project Science, Specifications, and Proposals,}
  \doarXiv{0910.2935}

\bibitem[{M. {Cadoni} {et~al.}(2026){Cadoni}, {de Lima}, {Pitzalis},
  {Rodrigues}, \& {Sanna}}]{CadoniDelima26}
{Cadoni}, M., {de Lima}, L., {Pitzalis}, M., {Rodrigues}, D.~C., \& {Sanna},
  A.~P. 2026, \bibinfo{title}{{Cosmologically Coupled Black Holes with Regular
  Horizons},} arXiv e-prints, arXiv:2601.03296,
  \dodoi{10.48550/arXiv.2601.03296}

\bibitem[{M. {Cadoni} {et~al.}(2024{\natexlab{a}}){Cadoni}, {Murgia},
  {Pitzalis}, \& {Sanna}}]{CadoniMurgia24}
{Cadoni}, M., {Murgia}, R., {Pitzalis}, M., \& {Sanna}, A.~P.
  2024{\natexlab{a}}, \bibinfo{title}{{Quasi-local masses and cosmological
  coupling of black holes and mimickers},} JCAP, 2024, 026,
  \dodoi{10.1088/1475-7516/2024/03/026}

\bibitem[{M. {Cadoni} {et~al.}(2024{\natexlab{b}}){Cadoni}, {Pitzalis},
  {Rodrigues}, \& {Sanna}}]{CadoniPitzalis24}
{Cadoni}, M., {Pitzalis}, M., {Rodrigues}, D.~C., \& {Sanna}, A.~P.
  2024{\natexlab{b}}, \bibinfo{title}{{Cosmological coupling of local
  gravitational systems},} JCAP, 2024, 045,
  \dodoi{10.1088/1475-7516/2024/08/045}

\bibitem[{M. {Cadoni} {et~al.}(2025){Cadoni}, {Pitzalis}, \&
  {Sanna}}]{CadoniPitzalis25}
{Cadoni}, M., {Pitzalis}, M., \& {Sanna}, A.~P. 2025, \bibinfo{title}{{Apparent
  horizons in cosmologically-embedded black holes},} JCAP, 2025, 051,
  \dodoi{10.1088/1475-7516/2025/02/051}

\bibitem[{M. {Cadoni} {et~al.}(2023){Cadoni}, {Sanna}, {Pitzalis}, {Banerjee},
  {Murgia}, {Hazra}, \& {Branchesi}}]{CadoniSanna23}
{Cadoni}, M., {Sanna}, A.~P., {Pitzalis}, M., {et~al.} 2023,
  \bibinfo{title}{{Cosmological coupling of nonsingular black holes},} JCAP,
  2023, 007, \dodoi{10.1088/1475-7516/2023/11/007}

\bibitem[{C. {Cain} {et~al.}(2025){Cain}, {Van Engelen}, {Croker}, {Kramer},
  {D'Aloisio}, \& {Lopez}}]{CainEngelen25}
{Cain}, C., {Van Engelen}, A., {Croker}, K.~S., {et~al.} 2025,
  \bibinfo{title}{{The Cosmic Microwave Background Optical Depth Constrains the
  Duration of Reionization},} \apjl, 987, L29, \dodoi{10.3847/2041-8213/ade141}

\bibitem[{M. {Calz{\`a}} {et~al.}(2024){Calz{\`a}}, {Gianesello}, {Rinaldi}, \&
  {Vagnozzi}}]{CalzaFrancesco25}
{Calz{\`a}}, M., {Gianesello}, F., {Rinaldi}, M., \& {Vagnozzi}, S. 2024,
  \bibinfo{title}{{Implications of cosmologically coupled black holes for
  pulsar timing arrays: Implications of cosmologically coupled...},} Sci. Rep.,
  14, 31296, \dodoi{10.1038/s41598-024-82661-8}

\bibitem[{D. Chattopadhyay {et~al.}(2021)Chattopadhyay, Stevenson, Hurley,
  Bailes, \& Broekgaarden}]{SKAmodelling}
Chattopadhyay, D., Stevenson, S., Hurley, J.~R., Bailes, M., \& Broekgaarden,
  F. 2021, \bibinfo{title}{{Modelling neutron star‚Äìblack hole binaries:
  future pulsar surveys and gravitational wave detectors},} Monthly Notices of
  the Royal Astronomical Society, 504, 3682, \dodoi{10.1093/mnras/stab973}

\bibitem[{K.~S. {Croker} {et~al.}(2020){Croker}, {Nishimura}, \&
  {Farrah}}]{CrokerNishimura20}
{Croker}, K.~S., {Nishimura}, K.~A., \& {Farrah}, D. 2020,
  \bibinfo{title}{{Implications of Symmetry and Pressure in Friedmann
  Cosmology. II. Stellar Remnant Black Hole Mass Function},} \apj, 889, 115,
  \dodoi{10.3847/1538-4357/ab5aff}

\bibitem[{K.~S. {Croker} {et~al.}(2024){Croker}, {Tarl{\'e}}, {Ahlen},
  {Cartwright}, {Farrah}, {Fernandez}, \& {Windhorst}}]{CrokerTarle24}
{Croker}, K.~S., {Tarl{\'e}}, G., {Ahlen}, S.~P., {et~al.} 2024,
  \bibinfo{title}{{DESI dark energy time evolution is recovered by
  cosmologically coupled black holes},} JCAP, 2024, 094,
  \dodoi{10.1088/1475-7516/2024/10/094}

\bibitem[{K.~S. {Croker} {\&} J.~L. {Weiner}(2019){Croker} \&
  {Weiner}}]{CrokerWeiner19}
{Croker}, K.~S., \& {Weiner}, J.~L. 2019, \bibinfo{title}{{Implications of
  Symmetry and Pressure in Friedmann Cosmology. I. Formalism},} \apj, 882, 19,
  \dodoi{10.3847/1538-4357/ab32da}

\bibitem[{K.~S. {Croker} {et~al.}(2022){Croker}, {Weiner}, \&
  {Farrah}}]{CrokerWeiner22}
{Croker}, K.~S., {Weiner}, J.~L., \& {Farrah}, D. 2022,
  \bibinfo{title}{{Well-defined equations of motion without constraining
  external sources},} \prd, 105, 084042, \dodoi{10.1103/PhysRevD.105.084042}

\bibitem[{K.~S. {Croker} {et~al.}(2021){Croker}, {Zevin}, {Farrah},
  {Nishimura}, \& {Tarl{\'e}}}]{CrokerZevin21}
{Croker}, K.~S., {Zevin}, M., {Farrah}, D., {Nishimura}, K.~A., \& {Tarl{\'e}},
  G. 2021, \bibinfo{title}{{Cosmologically Coupled Compact Objects: A
  Single-parameter Model for LIGO-Virgo Mass and Redshift Distributions},}
  \apjl, 921, L22, \dodoi{10.3847/2041-8213/ac2fad}

\bibitem[{K.~S. Croker {et~al.}(2026)Croker, Cohon, Toth, Rinaldi, Fernandez,
  Weiner, Farrah, Tarlé, Niz, Cain, Rohlf, Aviles, Summers, Thiele, Windhorst,
  \& Grumski-Flores}]{CrokerCohon26}
Croker, K.~S., Cohon, J., Toth, V.~T., {et~al.} 2026, A Strong Cosmologically
  Coupled Mass Growth Solution in Linearized General Relativity,
  \doarXiv{2609.38303}

\bibitem[{T. {Damour} {\&} N. {Deruelle}(1985){Damour} \&
  {Deruelle}}]{DamourDeruelle85}
{Damour}, T., \& {Deruelle}, N. 1985, \bibinfo{title}{{General relativistic
  celestial mechanics of binary systems. I. The post-Newtonian motion.},}
  Annales de L'Institut Henri Poincare Section (A) Physique Theorique, 43, 107

\bibitem[{T. {Damour} {\&} N. {Deruelle}(1986){Damour} \&
  {Deruelle}}]{DamourDeruelle86}
{Damour}, T., \& {Deruelle}, N. 1986, \bibinfo{title}{{General relativistic
  celestial mechanics of binary systems. II. The post-Newtonian timing
  formula.},} Annales de L'Institut Henri Poincare Section (A) Physique
  Theorique, 44, 263

\bibitem[{T. {Damour} {\&} J.~H. {Taylor}(1991){Damour} \&
  {Taylor}}]{DamourTaylor91}
{Damour}, T., \& {Taylor}, J.~H. 1991, \bibinfo{title}{{On the Orbital Period
  Change of the Binary Pulsar PSR 1913+16},} \apj, 366, 501,
  \dodoi{10.1086/169585}

\bibitem[{T. {Damour} {\&} J.~H. {Taylor}(1992){Damour} \&
  {Taylor}}]{DamourTaylor92}
{Damour}, T., \& {Taylor}, J.~H. 1992, \bibinfo{title}{{Strong-field tests of
  relativistic gravity and binary pulsars},} \prd, 45, 1840,
  \dodoi{10.1103/PhysRevD.45.1840}

\bibitem[{L.~V. {Drummond} {et~al.}(2026){Drummond}, {Chatziioannou}, \&
  {Fonseca}}]{Drummond26}
{Drummond}, L.~V., {Chatziioannou}, K., \& {Fonseca}, E. 2026,
  \bibinfo{title}{{Observational selection effects on radio pulsars are minimal
  for masses, but significant for orbits and spins},} \doarXiv{2609.03157}

\bibitem[{I. {Dymnikova}(2002){Dymnikova}}]{Dymnikova02}
{Dymnikova}, I. 2002, \bibinfo{title}{{The cosmological term as a source of
  mass},} Classical and Quantum Gravity, 19, 725,
  \dodoi{10.1088/0264-9381/19/4/306}

\bibitem[{A. {Einstein} {\&} E.~G. {Straus}(1945){Einstein} \&
  {Straus}}]{EinsteinStrauss45}
{Einstein}, A., \& {Straus}, E.~G. 1945, \bibinfo{title}{{The Influence of the
  Expansion of Space on the Gravitation Fields Surrounding the Individual
  Stars},} Reviews of Modern Physics, 17, 120,
  \dodoi{10.1103/RevModPhys.17.120}

\bibitem[{W. {Elbers}(2025){Elbers}}]{Elbers25}
{Elbers}, W. 2025, \bibinfo{title}{{Rapid late-time reionization: constraints
  and cosmological implications},} arXiv e-prints, arXiv:2508.21069,
  \dodoi{10.48550/arXiv.2508.21069}

\bibitem[{I. {Esteban} {et~al.}(2024){Esteban}, {Gonzalez-Garcia}, {Maltoni},
  {Martinez-Soler}, {Pinheiro}, \& {Schwetz}}]{Esteban24}
{Esteban}, I., {Gonzalez-Garcia}, M.~C., {Maltoni}, M., {et~al.} 2024,
  \bibinfo{title}{{NuFit-6.0: updated global analysis of three-flavor neutrino
  oscillations},} JHEP, 2024, 216, \dodoi{10.1007/JHEP12(2024)216}

\bibitem[{V. {Faraoni} {\&} M. {Rinaldi}(2024){Faraoni} \&
  {Rinaldi}}]{FaraoniRinaldi24}
{Faraoni}, V., \& {Rinaldi}, M. 2024, \bibinfo{title}{{Black hole event
  horizons are cosmologically coupled},} PRD, 110, 063553,
  \dodoi{10.1103/PhysRevD.110.063553}

\bibitem[{V. {Faraoni} {\&} M. {Rinaldi}(2026){Faraoni} \&
  {Rinaldi}}]{FaraoniRinaldi26}
{Faraoni}, V., \& {Rinaldi}, M. 2026, \bibinfo{title}{{Black holes become
  comoving with the universe},} arXiv e-prints, arXiv:2608.21222,
  \dodoi{10.48550/arXiv.2608.21222}

\bibitem[{D. {Farrah} {et~al.}(2023{\natexlab{a}}){Farrah}, {Croker}, {Zevin},
  {Tarl{\'e}}, {Faraoni}, {Petty}, {Afonso}, {Fernandez}, {Nishimura},
  {Pearson}, {Wang}, {Clements}, {Efstathiou}, {Hatziminaoglou}, {Lacy},
  {McPartland}, {Pitchford}, {Sakai}, \& {Weiner}}]{Farrah23b}
{Farrah}, D., {Croker}, K.~S., {Zevin}, M., {et~al.} 2023{\natexlab{a}},
  \bibinfo{title}{{Observational Evidence for Cosmological Coupling of Black
  Holes and its Implications for an Astrophysical Source of Dark Energy},}
  \apjl, 944, L31, \dodoi{10.3847/2041-8213/acb704}

\bibitem[{D. {Farrah} {et~al.}(2023{\natexlab{b}}){Farrah}, {Petty}, {Croker},
  {Tarl{\'e}}, {Zevin}, {Hatziminaoglou}, {Shankar}, {Wang}, {Clements},
  {Efstathiou}, {Lacy}, {Nishimura}, {Afonso}, {Pearson}, \&
  {Pitchford}}]{Farrah23a}
{Farrah}, D., {Petty}, S., {Croker}, K.~S., {et~al.} 2023{\natexlab{b}},
  \bibinfo{title}{{A Preferential Growth Channel for Supermassive Black Holes
  in Elliptical Galaxies at z {\ensuremath{\lesssim}} 2},} \apj, 943, 133,
  \dodoi{10.3847/1538-4357/acac2e}

\bibitem[{D. {Farrah} {et~al.}(2025){Farrah}, {Engholm}, {Hatziminaoglou},
  {Petty}, {Shankar}, {Efstathiou}, {Ejercito}, {Jones}, {Lacy}, {Lonsdale},
  {Pearson}, {Tarl{\'e}}, {Windhorst}, {Afonso}, {Clements}, {Croker}, \&
  {Pitchford}}]{Farrah25}
{Farrah}, D., {Engholm}, A., {Hatziminaoglou}, E., {et~al.} 2025,
  \bibinfo{title}{{The Assembly of Supermassive Black Holes at z < 1 in
  Early-type Galaxies from Scaling Relations},} \apj, 981, 71,
  \dodoi{10.3847/1538-4357/adb0c7}

\bibitem[{P.~C.~C. {Freire} {et~al.}(2017){Freire}, {Ridolfi}, {Kramer},
  {Jordan}, {Manchester}, {Torne}, {Sarkissian}, {Heinke}, {D'Amico}, {Camilo},
  {Lorimer}, \& {Lyne}}]{Freire17}
{Freire}, P.~C.~C., {Ridolfi}, A., {Kramer}, M., {et~al.} 2017,
  \bibinfo{title}{{Long-term observations of the pulsars in 47 Tucanae - II.
  Proper motions, accelerations and jerks},} \mnras, 471, 857,
  \dodoi{10.1093/mnras/stx1533}

\bibitem[{S.-J. {Gao} {\&} X.-D. {Li}(2023){Gao} \& {Li}}]{GaoLi2023}
{Gao}, S.-J., \& {Li}, X.-D. 2023, \bibinfo{title}{{Can Cosmologically Coupled
  Mass Growth of Black Holes Solve the Mass Gap Problem?},} \apj, 956, 128,
  \dodoi{10.3847/1538-4357/ace890}

\bibitem[{R. {Gaur} {\&} M. {Visser}(2024){Gaur} \& {Visser}}]{GaurVisser24}
{Gaur}, R., \& {Visser}, M. 2024, \bibinfo{title}{{Black holes embedded in FLRW
  cosmologies},} PRD, 110, 043529, \dodoi{10.1103/PhysRevD.110.043529}

\bibitem[{S. {Ghodla} {et~al.}(2023){Ghodla}, {Easther}, {Briel}, \&
  {Eldridge}}]{Ghodla23}
{Ghodla}, S., {Easther}, R., {Briel}, M.~M., \& {Eldridge}, J.~J. 2023,
  \bibinfo{title}{{Observational implications of cosmologically coupled black
  holes},} The Open Journal of Astrophysics, 6, 25,
  \dodoi{10.21105/astro.2306.08199}

\bibitem[{E.~B. {Gliner}(1966){Gliner}}]{Gliner66}
{Gliner}, E.~B. 1966, \bibinfo{title}{{Algebraic Properties of the
  Energy-momentum Tensor and Vacuum-like States o$^{+}$ Matter},} Sov. J. Exp.
  Theor. Phys., 22, 378

\bibitem[{H. {Goldstein}(1950){Goldstein}}]{Goldstein2nd}
{Goldstein}, H. 1950, {Classical mechanics}

\bibitem[{{}{H0DN Collaboration} {et~al.}(2026){H0DN Collaboration},
  {Casertano}, {Anand}, {Anderson}, {Beaton}, {Bhardwaj}, {Blakeslee},
  {Boubel}, {Breuval}, {Brout}, \& et~al.}]{Casertano26}
{H0DN Collaboration}, {Casertano}, S., {Anand}, G., {et~al.} 2026,
  \bibinfo{title}{{The Local Distance Network: A community consensus report on
  the measurement of the Hubble constant at ̃1\% precision},} \aap, 708, A166,
  \dodoi{10.1051/0004-6361/202557993}

\bibitem[{W.~E. {Harris}(1996){Harris}}]{Harris1996}
{Harris}, W.~E. 1996, \bibinfo{title}{{A Catalog of Parameters for Globular
  Clusters in the Milky Way},} \aj, 112, 1487, \dodoi{10.1086/118116}

\bibitem[{W.~E. {Harris}(2010){Harris}}]{Harris2010}
{Harris}, W.~E. 2010, \bibinfo{title}{{A New Catalog of Globular Clusters in
  the Milky Way},} arXiv e-prints, arXiv:1012.3224,
  \dodoi{10.48550/arXiv.1012.3224}

\bibitem[{A.~M. {Holgado} {\&} P.~M. {Ricker}(2019){Holgado} \&
  {Ricker}}]{HolgadoRicker19}
{Holgado}, A.~M., \& {Ricker}, P.~M. 2019, \bibinfo{title}{{Gravitational
  Radiation from Close Binaries with Time-varying Masses},} \apj, 882, 39,
  \dodoi{10.3847/1538-4357/ab3293}

\bibitem[{Y. {Kageura} {et~al.}(2026{\natexlab{a}}){Kageura}, {Ouchi},
  {Naokawa}, {Umeda}, {Matsumoto}, {Harikane}, {Nakane}, \& {Thai}}]{Kageura26}
{Kageura}, Y., {Ouchi}, M., {Naokawa}, F., {et~al.} 2026{\natexlab{a}},
  \bibinfo{title}{{A New Constraint on the Optical Depth from the Reionization
  History Independent of Cosmic Microwave Background Large-scale E-mode
  Polarization},} \apj, 1009, 42, \dodoi{10.3847/1538-4357/ae92fa}

\bibitem[{E.~F. {Keane} {et~al.}(2025){Keane}, {Graber}, {Levin}, {Tan},
  {Johnson}, {Ng}, {Pardo-Araujo}, {Ronchi}, {Vohl}, \& {Xue}}]{Keane25}
{Keane}, E.~F., {Graber}, V., {Levin}, L., {et~al.} 2025, \bibinfo{title}{{A
  Square Kilometre Array Pulsar Census},} The Open Journal of Astrophysics, 8,
  54256, \dodoi{10.33232/001c.154256}

\bibitem[{S.~A. {Klioner}(2016){Klioner}}]{Klioner16}
{Klioner}, S.~A. 2016, \bibinfo{title}{{Basic Celestial Mechanics},} arXiv
  e-prints, arXiv:1609.00915, \dodoi{10.48550/arXiv.1609.00915}

\bibitem[{F. {Kottler}(1918){Kottler}}]{Kottler18}
{Kottler}, F. 1918, \bibinfo{title}{{{\"U}ber die physikalischen Grundlagen der
  Einsteinschen Gravitationstheorie},} Annalen der Physik, 361, 401,
  \dodoi{10.1002/andp.19183611402}

\bibitem[{M. {Kramer} {et~al.}(2021){Kramer}, {Stairs}, {Manchester}, {Wex},
  {Deller}, {Coles}, {Ali}, {Burgay}, {Camilo}, {Cognard}, {Damour},
  {Desvignes}, {Ferdman}, {Freire}, {Grondin}, {Guillemot}, {Hobbs}, {Janssen},
  {Karuppusamy}, {Lorimer}, {Lyne}, {McKee}, {McLaughlin}, {M{\"u}nch},
  {Perera}, {Pol}, {Possenti}, {Sarkissian}, {Stappers}, \&
  {Theureau}}]{KramerStairs21}
{Kramer}, M., {Stairs}, I.~H., {Manchester}, R.~N., {et~al.} 2021,
  \bibinfo{title}{{Strong-Field Gravity Tests with the Double Pulsar},}
  Physical Review X, 11, 041050, \dodoi{10.1103/PhysRevX.11.041050}

\bibitem[{M. {Lacy} {et~al.}(2023){Lacy}, {Engholm}, {Farrah}, \&
  {Ejercito}}]{Lacy23}
{Lacy}, M., {Engholm}, A., {Farrah}, D., \& {Ejercito}, K. 2023,
  \bibinfo{title}{{Constraints on cosmological coupling from the accretion
  history of supermassive black holes},} Accepted at \apj, arXiv:2312.12344.
\newblock \doarXiv{2312.12344}

\bibitem[{L.~D. {Landau} {\&} E.~M. {Lifshitz}(1969){Landau} \&
  {Lifshitz}}]{LandauMechanics}
{Landau}, L.~D., \& {Lifshitz}, E.~M. 1969, {Mechanics}

\bibitem[{M. {Libralato} {et~al.}(2022){Libralato}, {Bellini}, {Vesperini},
  {Piotto}, {Milone}, {van der Marel}, {Anderson}, {Aparicio}, {Barbuy},
  {Bedin}, {Borsato}, {Cassisi}, {Dalessandro}, {Ferraro}, {King}, {Lanzoni},
  {Nardiello}, {Ortolani}, {Sarajedini}, \& {Sohn}}]{Libralato2022}
{Libralato}, M., {Bellini}, A., {Vesperini}, E., {et~al.} 2022,
  \bibinfo{title}{{The Hubble Space Telescope UV Legacy Survey of Galactic
  Globular Clusters. XXIII. Proper-motion Catalogs and Internal Kinematics},}
  \apj, 934, 150, \dodoi{10.3847/1538-4357/ac7727}

\bibitem[{D.~R. {Lorimer}(2008){Lorimer}}]{Lorimer08}
{Lorimer}, D.~R. 2008, \bibinfo{title}{{Binary and Millisecond Pulsars},}
  Living Reviews in Relativity, 11, 8, \dodoi{10.12942/lrr-2008-8}

\bibitem[{P.~O. {Mazur} {\&} E. {Mottola}(2015){Mazur} \&
  {Mottola}}]{MazurMottola15}
{Mazur}, P.~O., \& {Mottola}, E. 2015, \bibinfo{title}{{Surface tension and
  negative pressure interior of a non-singular {\textquoteleft}black
  hole{\textquoteright}},} Classical and Quantum Gravity, 32, 215024,
  \dodoi{10.1088/0264-9381/32/21/215024}

\bibitem[{P.~J. {McMillan}(2017){McMillan}}]{McMillan2017}
{McMillan}, P.~J. 2017, \bibinfo{title}{{The mass distribution and
  gravitational potential of the Milky Way},} \mnras, 465, 76,
  \dodoi{10.1093/mnras/stw2759}

\bibitem[{G.~C. {McVittie}(1933){McVittie}}]{McVittie33}
{McVittie}, G.~C. 1933, \bibinfo{title}{{The mass-particle in an expanding
  universe},} \mnras, 93, 325, \dodoi{10.1093/mnras/93.5.325}

\bibitem[{J.~C.~A. {Miller-Jones} {et~al.}(2021){Miller-Jones}, {Bahramian},
  {Orosz}, {Mandel}, {Gou}, {Maccarone}, {Neijssel}, {Zhao},
  {Zi{\'o}{\l}kowski}, {Reid}, \& et~al.}]{MillerJones21}
{Miller-Jones}, J. C.~A., {Bahramian}, A., {Orosz}, J.~A., {et~al.} 2021,
  \bibinfo{title}{{Cygnus X-1 contains a 21-solar mass black
  hole{\textemdash}Implications for massive star winds},} Science, 371, 1046,
  \dodoi{10.1126/science.abb3363}

\bibitem[{E. {Mlinar} {\&} T. {Zwitter}(2024){Mlinar} \&
  {Zwitter}}]{mlinarzwitter24}
{Mlinar}, E., \& {Zwitter}, T. 2024, \bibinfo{title}{{Determining cosmological
  growth parameter for stellar - mass black holes},} \mnras,
  \dodoi{10.1093/mnras/stae059}

\bibitem[{P.~C. {Peters}(1964){Peters}}]{Peters64}
{Peters}, P.~C. 1964, \bibinfo{title}{{Gravitational Radiation and the Motion
  of Two Point Masses},} Physical Review, 136, 1224,
  \dodoi{10.1103/PhysRev.136.B1224}

\bibitem[{{}{Planck Collaboration} {et~al.}(2020){Planck Collaboration},
  {Aghanim}, {Akrami}, {Arroja}, {Ashdown}, {Aumont}, {Baccigalupi},
  {Ballardini}, {Banday}, {Barreiro}, {Bartolo}, {Basak}, {Battye}, {Benabed},
  {Bernard}, {Bersanelli}, {Bielewicz}, {Bock}, {Bond}, {Borrill}, {Bouchet},
  {Boulanger}, {Bucher}, {Burigana}, {Butler}, {Calabrese}, {Cardoso},
  {Carron}, {Casaponsa}, {Challinor}, {Chiang}, {Colombo}, {Combet},
  {Contreras}, {Crill}, {Cuttaia}, {de Bernardis}, {de Zotti}, {Delabrouille},
  {Delouis}, {D{\'e}sert}, {Di Valentino}, {Dickinson}, {Diego}, {Donzelli},
  {Dor{\'e}}, {Douspis}, {Ducout}, {Dupac}, {Efstathiou}, {Elsner},
  {En{\ss}lin}, {Eriksen}, {Falgarone}, {Fantaye}, {Fergusson},
  {Fernandez-Cobos}, {Finelli}, {Forastieri}, {Frailis}, {Franceschi},
  {Frolov}, {Galeotta}, {Galli}, {Ganga}, {G{\'e}nova-Santos}, {Gerbino},
  {Ghosh}, {Gonz{\'a}lez-Nuevo}, {G{\'o}rski}, {Gratton}, {Gruppuso},
  {Gudmundsson}, {Hamann}, {Handley}, {Hansen}, {Helou}, {Herranz},
  {Hildebrandt}, {Hivon}, {Huang}, {Jaffe}, {Jones}, {Karakci}, {Keih{\"a}nen},
  {Keskitalo}, {Kiiveri}, {Kim}, {Kisner}, {Knox}, {Krachmalnicoff}, {Kunz},
  {Kurki-Suonio}, {Lagache}, {Lamarre}, {Langer}, {Lasenby}, {Lattanzi},
  {Lawrence}, {Le Jeune}, {Leahy}, {Lesgourgues}, {Levrier}, {Lewis},
  {Liguori}, {Lilje}, {Lilley}, {Lindholm}, {L{\'o}pez-Caniego}, {Lubin}, {Ma},
  {Mac{\'\i}as-P{\'e}rez}, {Maggio}, {Maino}, {Mandolesi}, {Mangilli},
  {Marcos-Caballero}, {Maris}, {Martin}, {Martinelli},
  {Mart{\'\i}nez-Gonz{\'a}lez}, {Matarrese}, {Mauri}, {McEwen}, {Meerburg},
  {Meinhold}, {Melchiorri}, {Mennella}, {Migliaccio}, {Millea}, {Mitra},
  {Miville-Desch{\^e}nes}, {Molinari}, {Moneti}, {Montier}, {Morgante}, {Moss},
  {Mottet}, {M{\"u}nchmeyer}, {Natoli}, {N{\o}rgaard-Nielsen}, {Oxborrow},
  {Pagano}, {Paoletti}, {Partridge}, {Patanchon}, {Pearson}, {Peel}, {Peiris},
  {Perrotta}, {Pettorino}, {Piacentini}, {Polastri}, {Polenta}, {Puget},
  {Rachen}, {Reinecke}, {Remazeilles}, {Renault}, {Renzi}, {Rocha}, {Rosset},
  {Roudier}, {Rubi{\~n}o-Mart{\'\i}n}, {Ruiz-Granados}, {Salvati}, {Sandri},
  {Savelainen}, {Scott}, {Shellard}, {Shiraishi}, {Sirignano}, {Sirri},
  {Spencer}, {Sunyaev}, {Suur-Uski}, {Tauber}, {Tavagnacco}, {Tenti},
  {Terenzi}, {Toffolatti}, {Tomasi}, {Trombetti}, {Valiviita}, {Van Tent},
  {Vibert}, {Vielva}, {Villa}, {Vittorio}, {Wandelt}, {Wehus}, {White},
  {White}, {Zacchei}, \& {Zonca}}]{PlanckDipole}
{Planck Collaboration}, {Aghanim}, N., {Akrami}, Y., {et~al.} 2020,
  \bibinfo{title}{{Planck 2018 results. I. Overview and the cosmological legacy
  of Planck},} \aap, 641, A1, \dodoi{10.1051/0004-6361/201833880}

\bibitem[{C.~L. {Rodriguez}(2023){Rodriguez}}]{Rodriguez23}
{Rodriguez}, C.~L. 2023, \bibinfo{title}{{Constraints on the Cosmological
  Coupling of Black Holes from the Globular Cluster NGC 3201},} \apjl, 947,
  L12, \dodoi{10.3847/2041-8213/acc9b6}

\bibitem[{G. {Sch{\"a}fer} {\&} P. {Jaranowski}(2024){Sch{\"a}fer} \&
  {Jaranowski}}]{SchaferJaranowski24}
{Sch{\"a}fer}, G., \& {Jaranowski}, P. 2024, \bibinfo{title}{{Hamiltonian
  formulation of general relativity and post-Newtonian dynamics of compact
  binaries},} Living Reviews in Relativity, 27, 2,
  \dodoi{10.1007/s41114-024-00048-7}

\bibitem[{P. {Shah} {et~al.}(2025){Shah}, {Davis}, {Vincenzi},
  {et~al.}}]{Shah25}
{Shah}, P., {Davis}, T.~M., {Vincenzi}, M., {et~al.} 2025 \mnras, 536, 946,
  \dodoi{10.1093/mnras/stae2614}

\bibitem[{B.~W. {Stappers} {et~al.}(2018){Stappers}, {Keane}, {Possenti}, \&
  {Stairs}}]{SKAprospects}
{Stappers}, B.~W., {Keane}, E.~F., {Possenti}, A., \& {Stairs}, I.~H. 2018,
  \bibinfo{title}{{The prospects of pulsar timing with new-generation radio
  telescopes and the Square Kilometre Array},} Philosophical Transactions of
  the Royal Society A, 376, Issue 2120, \dodoi{10.1098/rsta.2017.0293}

\bibitem[{J.~H. {Taylor} {et~al.}(1979){Taylor}, {Fowler}, \&
  {McCulloch}}]{TaylorFowler79}
{Taylor}, J.~H., {Fowler}, L.~A., \& {McCulloch}, P.~M. 1979,
  \bibinfo{title}{{Measurements of general relativistic effects in the binary
  pulsar PSR1913 + 16},} \nat, 277, 437, \dodoi{10.1038/277437a0}

\bibitem[{M. {Tristram} {et~al.}(2024){Tristram}, {Banday}, {Douspis},
  {Garrido}, {G{\'o}rski}, {Henrot-Versill{\'e}}, {Hergt}, {Ili{\'c}},
  {Keskitalo}, {Lagache}, \& et~al.}]{PlanckPR4}
{Tristram}, M., {Banday}, A.~J., {Douspis}, M., {et~al.} 2024,
  \bibinfo{title}{{Cosmological parameters derived from the final Planck data
  release (PR4)},} \aap, 682, A37, \dodoi{10.1051/0004-6361/202348015}

\bibitem[{C.~M. {Will}(2018){Will}}]{TEGP18}
{Will}, C.~M. 2018, {Theory and Experiment in Gravitational Physics},
  \dodoi{10.1017/9781316338612}

\bibitem[{N. {Yunes} {et~al.}(2025){Yunes}, {Siemens}, \&
  {Yagi}}]{YunesSiemens25}
{Yunes}, N., {Siemens}, X., \& {Yagi}, K. 2025,
  \bibinfo{title}{{Gravitational-wave tests of general relativity with
  ground-based detectors and pulsar-timing arrays},} Living Reviews in
  Relativity, 28, 3, \dodoi{10.1007/s41114-024-00054-9}

\end{thebibliography}
\bibliographystyle{aasjournalv7.1}
\end{document}